\documentclass[11pt,english,reqno]{article}

\usepackage[latin9]{inputenc}
\usepackage{geometry}
\usepackage{enumerate}
\usepackage{color}
\usepackage{babel}
\usepackage{mathrsfs}
\usepackage{bm}
\usepackage{amsmath}
\usepackage{amssymb}
\usepackage{graphicx}
\usepackage{setspace}
\usepackage{esint}
\usepackage[unicode=true,pdfusetitle,
 bookmarks=true,bookmarksnumbered=false,
 bookmarksopen=true,bookmarksopenlevel=2,
 breaklinks=false,pdfborder={0 0 0},backref=false,colorlinks=true,linktoc=all]
 {hyperref}
\makeatletter

\numberwithin{equation}{section}
\usepackage{babel}
\usepackage[usenames,dvipsnames]{xcolor}
\hypersetup{
    linkcolor=BlueViolet,
    citecolor=Mahogany,
    urlcolor=ForestGreen
}
\makeatother
\usepackage[sf,bf]{caption}
\usepackage{soul}
\usepackage{cite}
\usepackage{epstopdf}

\title{\bf{Geoids in General Relativity: Geoid Quasilocal Frames}}

\date{\today}

\usepackage{authblk}

\author[1,2]{Marius Oltean\thanks{\texttt{\textcolor{blue}{\href{mailto:moltean@cnrs-orleans.fr}{moltean@cnrs-orleans.fr}}}}}

\author[3]{Richard J. Epp\thanks{\texttt{\textcolor{blue}{\href{mailto:rjepp@uwaterloo.ca}{rjepp@uwaterloo.ca}}}}}

\author[3]{Paul L. McGrath\thanks{\texttt{\textcolor{blue}{\href{mailto:pmcgrath@wlu.ca}{pmcgrath@uwaterloo.ca}}}}}

\author[3,4]{Robert B. Mann\thanks{\texttt{\textcolor{blue}{\href{mailto:rbmann@uwaterloo.ca}{rbmann@uwaterloo.ca}}}}}

\affil[1]{\small{\it{Institute of Space Sciences (CSIC-IEEC), Campus Universitat Aut\`{o}noma de Barcelona\\

Carrer de Can Magrans s/n, 08193 Cerdanyola del Vall\`{e}s, Barcelona, Spain
\vspace*{0.2cm}}}}

\affil[2]{\small{\it{Laboratoire de Physique et Chimie de l'Environnement et de l'Espace\\

Centre National de la Recherche Scientifique, Universit\'{e} d'Orl\'{e}ans\\

3A Avenue de la Recherche Scientifique, 45071 Orl\'{e}ans, France
\vspace*{0.2cm}}}}

\affil[3]{\it{Department of Physics and Astronomy, University of Waterloo\\

200 University Avenue West, Waterloo, Ontario N2L 3G1, Canada
\vspace*{0.2cm}}}

\affil[4]{\it{Perimeter Institute for Theoretical Physics\\

31 Caroline Street North, Waterloo, Ontario N2L 2Y5, Canada}}

\begin{document}
\maketitle


\begin{abstract}

\normalsize{We develop, in the context of general relativity, the notion of a geoid -- a surface of constant ``gravitational potential". In particular, we show how this idea naturally emerges as a specific choice of a previously proposed, more general and operationally useful construction called a quasilocal frame -- that is, a choice of a two-parameter family of timelike worldlines comprising the worldtube boundary of the history of a finite spatial volume. We study the geometric properties of these geoid quasilocal frames, and construct solutions for them in some simple spacetimes. We then compare these results -- focusing on the computationally tractable scenario of a non-rotating body with a quadrupole perturbation -- against their counterparts in Newtonian gravity (the setting for current applications of the geoid), and we compute general-relativistic corrections to some measurable geometric quantities.}

~

~

~

~

~

~

~

\end{abstract}
\thispagestyle{empty}

\pagebreak{}
\noindent \begin{center}
\rule{1\columnwidth}{1pt}
\par\end{center}
\tableofcontents{}
\noindent \begin{center}
\rule{1\columnwidth}{1pt}
\par\end{center}

\section{Introduction}
\label{sec:1}

Recent satellite missions of the European Space Agency, including
GRACE (Gravity Recovery and Climate Experiment) \cite{GRACE} and
GOCE (Gravity Field and Steady-State Ocean Circulation Explorer) \cite{GOCE},
have been able to produce high-resolution maps of the Earth's geoid, a surface with a particular gravity potential value \cite{EGM98,Moritz,NOAA}. More specifically, it is a very particular equipotential surface which most closely approximates an idealized mean global ocean surface (i.e. the shape the oceans would take under the influence of gravity alone). An understanding of the geoid is salient for a wide
variety of theoretical and technological applications in the realm
of gravitational physics. Foremost among them, knowledge of the geoid
map is used by all GPS satellites in order to calibrate their height
measurements \cite{GPS,NOAA}. In particular, the geoid model is what is used to translate mathematical heights (measurements relative to a reference ellipsoid obtained from GPS) into positions of real-world significance (global elevations relative to a meaningful physical reference surface, i.e. the sea level) \cite{NOAA}. An accurate determination of the Earth's
geoid is also of importance for studying geophysical processes, climate
patterns, oceanic tides and related phenomena.

Historically, the theory of the geoid has
been extensively -- and, until very recently, exclusively -- explored
in the context of Newtonian gravity (NG). The first -- and, up until
the present work, only -- analysis of the geoid in the context of
general relativity (GR)  appeared in \cite{Kopeikin:2014afa}.
There, the authors argue for and derive, in a particular case, the
PDEs that a geoid should satisfy assuming a certain simple model of
the Earth.

The scope of the present work is two-fold. On the one hand, we feel
that the geoid concept in GR, rather than a mere expedient imposition
of certain conditions on the metric functions in order to achieve
semblance wih NG (as is the approach of \cite{Kopeikin:2014afa}),
arises as a particular manifestation of a much broader and richer
idea known as a \textsl{quasilocal frame} \cite{Epp:2008kk,Epp:2013xza}.
Developed over the course of the past few years, this concept has
proven to be a very operationally practical and geometrically natural
way of describing extended systems in GR. The basic idea is to work
with finite volumes of space whose dynamics are such that their spatial
boundaries obey certain conditions or constraints as they evolve in
time. In work done up to now, the imposed constraints were those of
\textsl{rigidity} -- for the simple reason that these manifestly simplify and
naturally lend themselves to a variety of both theoretical as well as practical
problems. They include, in the case of the former, the formulation
of conservation laws in GR without the traditional need to appeal
to spacetime symmetries \cite{McGrath:2012db,Epp:2013hua}; and in
the case of the latter, the calculation of tidal interactions in solar
system dynamics in a post-Newtonian setting \cite{McGrath:2013pea}. Yet, as will be shown, there
are enough degrees of freedom in choosing these boundary constraints
such that, instead of making them rigid, they can be chosen to actually describe
a geoid -- and that these two possible choices represent naturally
complementary views of a quasilocal approach to GR.

A second objective of this work is to explicitly construct geoid solutions
in GR and to exemplify their disagreement with the Newtonian theory.
Thus, we will argue that a general-relativistic treatment of the geoid
may be important not only from a foundational point of view, but also
for future and improved applications.

We structure this paper as follows. In Section \ref{sec:QF}, we provide a self-contained overview of the formalism of quasilocal frames, and in Section \ref{sec:ConsLaws}, of conservation laws arising therefrom, \textsl{in general}, i.e. prior to imposing any specializing conditions or constraints thereon. In Section \ref{sec:GQF}, we show how these conditions can be chosen so that quasilocal frames describe geoids -- hence, geoid quasilocal frames. Then, in Section \ref{sec:Solns}, we construct solutions for them in some stationary axisymmetric spacetimes, and in Section \ref{sec:NGvGR}, we focus on one particular class of such solutions -- non-rotating bodies with a quadrupole perturbation -- and compare them with their analogues in NG. Finally, in Section \ref{sec:Concl} we offer some concluding remarks.

\section{Quasilocal Frames}
\label{sec:QF}

\subsection{Setup}
\label{sub:Spacetime-Decomposition}

Let $\mathscr{M}$ be a 4-manifold with a Lorentzian metric
$g_{ab}$ of signature $(-,+,+,+)$ and
a metric-compatible derivative operator $\nabla$.
For any type $(k,l)$ tensor $A\in\mathscr{T}_{l}^{k}\left(\mathscr{M}\right)$,
we use \textsl{Roman} indices to denote its components $A^{a_{1}\cdots a_{k}}\,_{b_{1}\cdots b_{l}}$
in some coordinate basis $\left\{ x^{a}\right\} $.

\begin{figure}
\begin{centering}
\includegraphics[scale=0.4]{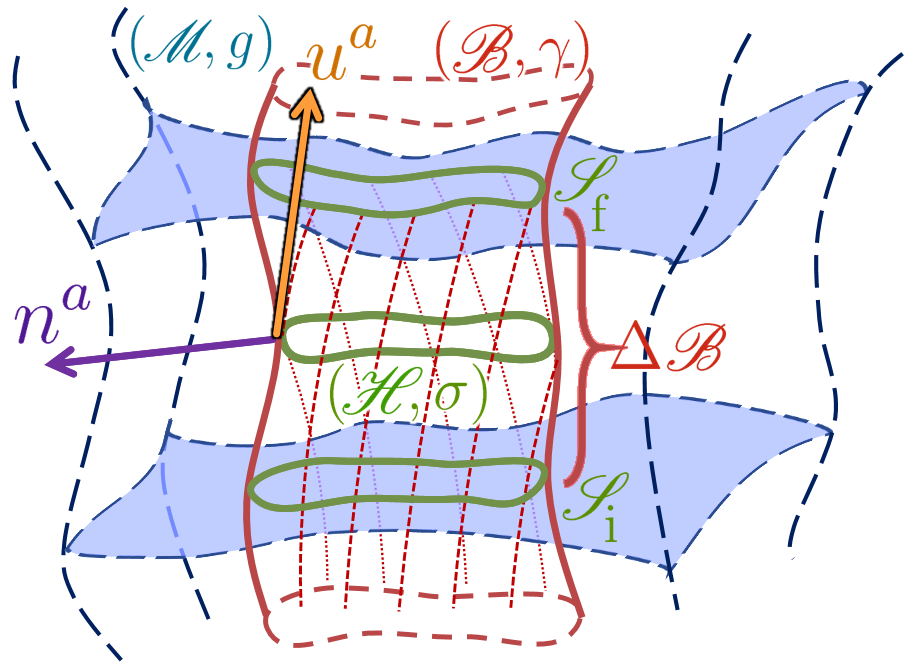}
\par\end{centering}
\protect\caption{\label{fig:QF}Definition of a quasilocal frame $\left(\mathscr{B},\gamma\right)$, illustrated for ease of visualization in the situation where the orthogonal subspaces $\left(\mathscr{H},\sigma\right)$ are closed two-surfaces.}
\end{figure}

Let $\mathscr{B}\subset\mathscr{M}$ be an embedded 3-manifold representing a two-parameter
family of timelike curves in $\mathscr{M}$, such that $\mathscr{B}\simeq\mathbf{R}\times\mathbf{S}^{2}$
is the worldtube boundary of the history of a finite spatial volume. This is what is referred to as a quasilocal frame (QF); see Figure \ref{fig:QF}.
Let $u^{a}$ denote the the future-directed unit vector field tangent
to this congruence. The metric on $\mathscr{M}$ induces on $\mathscr{B}$
a spacelike outward-pointing unit normal vector $n^{a}$ and a Lorentzian
metric of signature $(-,+,+)$ whose components in $\mathscr{M}$
are given by
\begin{equation}
\boxed{\gamma_{ab}=g_{ab}-n_{a}n_{b}.}
\label{eq:gamma_ab}
\end{equation}
Moreover, there is
an induced metric-compatible
(with respect to $\boldsymbol{\gamma}$) derivative operator $D$
on $\mathscr{B}$. Here, for any type $(k,l)$ tensor $\bar{A}\in\mathscr{T}_{l}^{k}\left(\mathscr{B}\right)$,
we use \textsl{Greek} indices to denote its components $\bar{A}^{\mu_{1}\cdots\mu_{k}}\,_{\nu_{1}\cdots\nu_{l}}$
in some coordinate basis $\left\{ \bar{x}^{\mu}\right\} $, and overbars
are used to make it clear that quantities belong to $\mathscr{B}$.

Finally, let $\mathscr{H}$ be the subspace of the tangent space of $\mathscr{B}$ that is orthogonal to $u^{a}$. Note that $\mathscr{H}$ need not be integrable into closed 2-dimensional hypersurfaces (with topology $\mathbf{S}^{2}$) since $u^{a}$ is in general not hypersurface orthogonal, though to make the setup easy to visualize, a representative example of such a closed 2-dimensional hypersurfaces is depicted in Figure \ref{fig:QF}.
Regardless, we always have a 2-dimensional Riemannian metric induced on $\mathscr{H}$ of signature $(+,+)$ with components in $\mathscr{M}$ given by 
\begin{equation}
\boxed{\sigma_{ab}=\gamma_{ab}+u_{a}u_{b}.}
\label{eq:sigma_ab}
\end{equation}
Analogously, there is an induced metric-compatible
(with respect to $\boldsymbol{\sigma}$) derivative operator $\mathfrak{D}$
on $\mathscr{H}$. Here, for any type $(k,l)$ tensor $\hat{A}\in\mathscr{T}_{l}^{k}\left(\mathscr{H}\right)$,
we use \textsl{Fraktur} indices to denote its components $\hat{A}^{\mathfrak{i}_{1}\cdots\mathfrak{i}_{k}}\,_{\mathfrak{j}_{1}\cdots\mathfrak{j}_{l}}$
in some coordinate basis $\{\hat{x}^{\mathfrak{i}}\}$, and overhats
are used to make it clear that quantities belong to $\mathscr{H}$.

In any given spacetime $\left(\mathscr{M},g\right)$,
the specification of a QF $\left(\mathscr{B},\gamma\right)$ lies entirely in the choice of the vector field
$u^{a}$ (from which $n^{a}$ and all induced quantities are then computable). This therefore affords us \textsl{three} degrees of freedom in how we construct a QF, associated with the three (a priori) independent
components of $u^{a}$ (whose total of four components are subject
to the normalization condition $\bm{u}\cdot\bm{u}=-1$). In other
words, we are free to pick any three mathematical conditions to fix (uniquely) the geometry
of a QF. In practice (as we shall see), it is physically more natural (as well as mathematically easier) to work with geometric quantities other than $\bm{u}$ itself to achieve this.

\subsection{Induced Geometry}

Let $\mathbb{J}^{I}\,_{J}=\partial x^{I}/\partial x^{J}$,
where $I$ and $J$ can be Roman, Greek or Fraktur (so long as $J$ is of a type
that refers to a dimension lower than or equal to $I$). If they are of
the \textsl{same} type, $\mathbb{J}$ accomplishes a coordinate transformation within
the same manifold (i.e. it is just a Jacobian); if they are of \textsl{different} types, we use $\mathbb{J}$
as a means to compute the coordinate components of tensors induced on the corresponding lower-dimensional submanifold, viz. $\forall A\in\mathscr{T}_{l}^{0}\left(\mathscr{M}\right)$
we obtain its projections $\bar{A}\in\mathscr{T}_{l}^{0}\left(\mathscr{B}\right)$
and $\hat{A}\in\mathscr{T}_{l}^{0}\left(\mathscr{H}\right)$, respectively,
by $\bar{A}_{\mu_{1}\cdots\mu_{l}}=\left(\prod_{j=1}^{l}\mathbb{J}^{a_{j}}\,_{\mu_{j}}\right)A_{a_{1}\cdots a_{l}}$
and $\hat{A}_{\mathfrak{i}_{1}\cdots\mathfrak{i}_{l}}=\left(\prod_{j=1}^{l}\mathbb{J}^{\mu_{j}}\,_{\mathfrak{i}_{j}}\right)\bar{A}_{\mu_{1}\cdots\mu_{l}}$.

Moreover, if we define on $\mathscr{B}$ a time function $t$, we
can choose coordinates $\{\bar{x}^{\mu}\}=\{t,\bar{x}^{\mathfrak{i}}\}$
so that $\bar{x}^{\mathfrak{i}}=\hat{x}^{\mathfrak{i}}$ and set $\bar{u}^{\mu}=\left(1/N\right)\bar{\delta}^{\mu}\,_{0}$
where $N$ is the lapse function of $g_{ab}$. Then, we find that
the metric on $\mathscr{B}$ has adapted coordinate components: 
\begin{equation}
\bar{\gamma}_{\mu\nu}=\left(\begin{array}{cc}
-N^{2} & N\hat{u}_{\mathfrak{j}}\\
N\hat{u}_{\mathfrak{i}} & \hat{\sigma}_{\mathfrak{ij}}-\hat{u}_{\mathfrak{i}}\hat{u}_{\mathfrak{j}}
\end{array}\right).\label{eq:gamma}
\end{equation}

Additionally, we denote the extrinsic curvature of $\mathscr{B}$
by $\Theta_{ab}=\gamma_{ac}\nabla^{c}n_{b},$ and $\Theta=\Theta_{a}\,^{a}$
is its trace. From this, the canonical momentum $\Pi_{ab}=\Theta_{ab}-\Theta\gamma_{ab}$
of $\mathscr{B}$ can be computed. Moreover, we also define the quantity $\kappa_{ab}=\sigma_{ac}\nabla^{c}n_{b}$ which will commonly appear in our computations, and $\kappa=\kappa_{a}\,^{a}$ is its trace.

\subsection{Congruence Properties}

A natural quantity for describing the time development of our congruence
is the tensor field
\begin{equation}
\boxed{\theta_{ab}=\sigma_{ac}\sigma_{bd}\nabla^{c}u^{d}=\frac{1}{2}\theta\sigma_{ab}+\theta_{\langle ab\rangle}+\theta_{\left[ab\right]},}\label{eq:theta}
\end{equation}
known as the \textsl{strain rate tensor}. In the last equality it
is written in the usual decomposition\footnote{See, e.g., Chapter 2 of \cite{Poisson} for a detailed discussion.}  into, respectively: 
its trace part, proportional to $\theta=\sigma^{ab}\theta_{ab}=\theta_{a}\,^{a}$, describing \textsl{expansion};
its symmetric trace-free part, $\theta_{\langle ab\rangle}=\theta_{\left(ab\right)}-\frac{1}{2}\theta\sigma_{ab}$, describing \textsl{shear};
and its totally antisymmetric part, $\theta_{\left[ab\right]}$, describing \textsl{rotation}.
They vanish if, respectively, the congruence does not change size,
does not change shape, or does not rotate.

Let us now see how (\ref{eq:theta}) can
help us study properties of the congruence. Let $a^{a}=\nabla_{\boldsymbol{u}}u^{a}$
denote the acceleration of the congruence, $\alpha^{a}=\sigma^{ab}a_{b}$
the acceleration tangential to $\mathscr{B}$, and $\aleph=\boldsymbol{n}\cdot\boldsymbol{a}=n_{a}a^{a}$
the normal component of the acceleration. Computing $u^{c}\nabla_{c}\theta_{ab}=\nabla_{\boldsymbol{u}}\theta_{ab}$
and taking its trace, we obtain a Raychaudhuri-type equation prescribing
the rate of change of $\theta$ with respect to proper time; the details
of the calculation are offered in Appendix \ref{sec:A}, and here
we state only the result: 
\begin{equation}
\frac{{\rm d}\theta}{{\rm d}\tau}=-\theta_{ab}\theta^{ba}-R_{ab}u^{a}u^{b}+\boldsymbol{D}\cdot\boldsymbol{\alpha}+\boldsymbol{\mathcal{P}}^{2}+\aleph\kappa+R_{acbd}n^{a}n^{b}u^{c}u^{d},\label{eq:Raychaudhuri}
\end{equation}
where the quantity $\mathcal{P}^{a}=\sigma^{ab}u^{c}\Pi_{bc}$ is
known as the quasilocal momentum, the physical meaning of which we
will explicate in greater detail in the following section.

The first two terms on the RHS of (\ref{eq:Raychaudhuri}) are the
familiar ones from the standard Raychaudhuri equation for timelike geodesics \cite{Poisson,Wald};
the rest are due to the fact that here we are considering a general congruence
with \textsl{arbitrary} (non-zero) acceleration, and that we have defined
the strain rate tensor (\ref{eq:theta}) not as the derivative (in $\mathscr{M}$)
of the tangent vector to a 3-parameter congruence (as is usually done \cite{Poisson,Wald}), but instead as
the projection thereof onto $\mathscr{H}$ for a 2-parameter congruence.

The utility of this is that it permits us to impose (some or all) conditions
on (\ref{eq:theta}) in order to define a QF (which is often much
more meaningful than, for example, directly specifying $\bm{u}$).
Moreover, (\ref{eq:Raychaudhuri}) provides us with a consistency
check once we construct our QFs (as we shall see in later sections)
and may in certain cases encode useful physical insight into the nature
of the solutions.

\section{Conservation Laws}
\label{sec:ConsLaws}

To see what QF conditions make sense and are practical to work with,
let us turn now to what has so far proven to be the principal utility
of this setup: namely, the construction and calculational implementation
of general-relativistic conservation laws \cite{McGrath:2012db,Epp:2013hua}.
The starting point for this is the well-known fact that the usual
(local) stress-energy-momentum (SEM) tensor of matter $T_{ab}$ bears
a number of manifest deficiencies when taken as the sole ingredient
for formulating conservation laws in dynamically curved spacetimes.
Broadly speaking, there are two reasons why: Local conservation laws
constructed from $T_{ab}$ cannot
\begin{enumerate}
\item be expressed, in general, purely in terms of boundary fluxes (which
is, intuitively, the mathematical form we should expect/demand of
conservation laws), but must include four-dimensional bulk integrals
(unless the spacetime posesses Killing vector fields, which in general
is not the case if we have nontrivial dynamics);
\item account for gravitational effects, i.e. they have nothing to say about
vacuum solutions (which may contain nontrivial fluxes of gravitational energy or momentum, e.g. from gravitational waves), and in any case, gravitational SEM is (unlike matter) \textsl{not}
localizable.
\end{enumerate}
A solution that alleviates both of these problems is to define and
work with a SEM tensor for \textsl{both matter and gravity}. The most
intuitive argument for how to achieve this \cite{Epp:2013hua} follows by recalling the basic definition
of the matter SEM tensor $T_{ab}$ itself: namely, it is twice the
variational derivative of the total action functional $S$ with respect
to the spacetime metric $g$ (which yields a \textsl{bulk} integral), if gravity is \textsl{not}
included in $S$. But if it \textsl{is}, and the Einstein equation
$G_{ab}=T_{ab}$ is assumed to hold everywhere in the bulk, then (with
the bulk integral vanishing) this variational derivative simply evaluates
to $\delta_{g}S=\frac{1}{2}\int{\rm d}\mathscr{B}(-\Pi^{ab})\delta\gamma_{ab}$, which is now
a \textsl{boundary} integral (over $\mathscr{B}$). This motivates taking $\tau_{ab}=-\Pi_{ab}$ as the total \textsl{matter plus gravity}
SEM tensor; it is also known as the
\textsl{Brown-York tensor} (after Brown and York, who first proposed this idea \cite{Brown-York:1993,Brown:2000dz}, but following instead an argument relying on a Hamilton-Jacobi analysis). It is \textsl{quasilocal} in the sense that it describes the boundary
(as opposed to volume) density of energy, momentum and stress -- obtained,
respectively, by decomposing $\tau_{ab}$ into: $\mathcal{E}=u^{a}u^{b}\tau_{ab}$,
$\mathcal{P}_{a}=-\sigma_{a}\,^{b}u^{c}\tau_{bc}$ and $\mathcal{S}_{ab}=-\sigma_{a}\,^{c}\sigma_{b}\,^{d}\tau_{cd}$.

Conservation laws using $\tau_{ab}$ have been explicitly constructed
(without the existence of Killing vectors, or any other extraneous
assumptions), applied successfully to simple situations (such as a
box accelerating through a uniform electric field, viz. the Bertotti-Robinson-type metric), and interpreted
as offering physically sensible and intuitive explanations for the
mechanisms behind energy and momentum transfer in curved spacetime
\cite{McGrath:2012db,Epp:2013hua}. For brevity, we here explicitly
write down only the energy conservation law. (The ones for momentum
and angular momentum are analogous, but a bit more involved.) It expresses
the change in total energy between an initial and final arbitrary
spatial volume, with spatial boundaries $\mathscr{S}_{{\rm i}}$ and
$\mathscr{S}_{{\rm f}}$ respectively (see Figure \ref{fig:QF}), as a flux through the worldtube
boundary $\Delta\mathscr{B}$ between them%
\footnote{If the observers are not at rest with respect to $\mathscr{S}_{{\rm i}}$
and $\mathscr{S}_{{\rm f}}$, the LHS actually involves an additional
shift term (omitted here for simplicity).%
}:
\begin{equation}
\intop_{\mathscr{S}_{{\rm f}}-\mathscr{S}_{{\rm i}}}{\!\!\!\rm d}\mathscr{H}\,\mathcal{E}=\intop_{\Delta\mathscr{B}}{\rm d}\mathscr{B}\left[T^{ab}n_{a}u_{b}-\left(\alpha_{a}\mathcal{P}^{a}-\mathcal{S}_{ab}\theta^{ab}\right)\right].\label{eq:EnergyConservationLaw}
\end{equation}

What (\ref{eq:EnergyConservationLaw}) is saying is that the change in
energy of a system (the LHS) occurs as a consequence of two processes (on the RHS): a \textsl{matter}
flux through $\Delta\mathscr{B}$, i.e. the $\int{\rm d}\mathscr{B}T^{ab}n_{a}u_{b}$
term, plus a purely geometrical flux, $\int{\rm d}\mathscr{B}\left(-\alpha_{a}\mathcal{P}^{a}+\mathcal{S}_{ab}\theta^{ab}\right)$,
which can be interpreted as representing a \textsl{gravitational}
energy flux through the boundary. The form of this flux provides
a natural way to identify the two complementary classes or ``gauges''
of QFs: ones which see gravitational energy flux as \textsl{momentum times
acceleration}, and ones which see it as \textsl{stress times strain.}

The first gauge, dubbed \textsl{rigid }quasilocal frames (RQFs), has
been explored extensively -- and exclusively -- prior to the
present work. It is defined by the three QF conditions 
\begin{equation}
\boxed{\theta=0=\theta_{\langle ab\rangle},}\label{eq:RQFconditions}
\end{equation}
i.e. the congruence does not expand and undergoes no shearing -- which
is another way of saying that observers remain at rest or ``rigid''
with respect to each other. In such QFs, the gravitational energy flux is $-\bm{\alpha}\cdot\bm{\mathcal{P}}$, which is simply the special relativistic rate of change of energy, $-\bm{a}\cdot \bm{p}$, of an object with four-momentum $\bm{p}$ as seen by an observer with four-acceleration $\bm{a}$, promoted to a general-relativistic energy flux passing through a rigid boundary containing the object: $\bm{a}$ becomes $\bm{\alpha}$, the acceleration of observers on the boundary, and $\bm{p}$ becomes $\bm{\mathcal{P}}$, the quasilocal momentum density measured by those observers (via general-relativistic frame dragging). An equivalence principle-based argument explaining why this must be so is given in \cite{McGrath:2012db,Epp:2013hua}.

The second gauge of QFs, wherein the $\bm{\alpha}\cdot\bm{\mathcal{P}}$ term
does not contribute and the gravitational energy flux is transmitted solely
through the effect of stress times strain, can be interpreted as describing
\textsl{geoids}, and we hence refer to these QFs as \textsl{geoid
quasilocal frames} (GQFs); we elaborate further in the following section.

\section{Geoid Quasilocal Frames}
\label{sec:GQF}

Geoids in NG are defined as equipotential surfaces,
i.e. the level sets of the Newtonian gravitational potential $V_{{\rm N}}$,
which is the solution (in $\mathbf{R}^{3}$) of the Poisson equation,
$\nabla^{2}V_{{\rm N}}=\rho$ (where $\rho$ denotes mass density).
To define a geoid in GR, which (unlike NG) does \textsl{not} rely
on the concept of a potential, we can instead take our cue from the expected motion
of observers associated therewith. In NG, acceleration is proportional
to $\nabla V_{{\rm N}}$ (whose direction is perpendicular to the
level surface), and so geoids can alternatively be regarded as surfaces
the normal vector to which always indicates the direction of observers'
acceleration. This \textsl{is} a condition that can, just as well,
be imposed in GR. In particular, it is achieved precisely in the form
of a QF where the acceleration of observers $a^{a}$
points only in the direction of the normal vector $n^{a}$. This is
the same as saying that there is no component of $a^{a}$ tangent
to $\mathscr{B}$, i.e. $\alpha^{a}=\sigma^{ab}a_{b}=0$. But this
comprises just \textsl{two} equations (since $\bm{a}\cdot \bm{u}=0$ holds automatically),
and we have \textsl{three} degrees of freedom at our disposal in specifying
a QF. Thus, as our third condition for geoids, we impose -- for reasons
we will explicate momentarily -- the requirement of zero expansion
of the congruence, i.e. $\theta=0$. To summarize, the GQF conditions
are:
\begin{equation}
\boxed{\theta=0=\alpha^{a}.}\label{eq:GQFconditions}
\end{equation}
These generalize the conditions proposed in \cite{Kopeikin:2014afa},
which use just an equivalent version of $\alpha^{a}=0$, and in fact one that
is only valid under the assumption of stationarity; our conditions (\ref{eq:GQFconditions}) are thus completely general and, moreover, the geometrical meaning of the geoid is more apparently elucidated in our formulation.

More conveniently for applications, it can be shown that (\ref{eq:GQFconditions}) may be rewritten as the following equivalent system of PDEs in the adapted coordinates:
\begin{align}
\alpha^{a}=0 & \Leftrightarrow0=\partial_{\mathfrak{i}}N+\dot{\hat{u}}_{\mathfrak{i}},\label{eq:GQFcondition1}\\
\theta=0 & \Leftrightarrow0=\hat{\sigma}^{\mathfrak{ij}}\dot{\hat{\sigma}}_{\mathfrak{ij}},\label{eq:GQFcondition2}
\end{align}
where overdots denote time differentiation.

The choice of the $\theta=0$ condition may be justified in a few
ways: 
\begin{itemize}
\item It is physically intuitive: For example, in a static spacetime, a
globally non-zero value of $\theta$ would describe a changing total surface
area, which is not what one might think of as a geoid. If, moreover,
we merely imposed that the total surface area be constant, but with
$\theta$ (potentially) positive on some parts of the geoid but negative
on others, the quasilocal observers would be dynamical relative to
each other (either converging toward each other on some parts and
diverging away on other parts, or moving radially outwards on some
parts and radially inwards on other parts), which is again a situation
that one would not think of as describing a geoid. Hence $\theta=0$
is a physically natural choice as a (local) GQF condition.
\item It reduces the gravitational energy flux in (\ref{eq:EnergyConservationLaw})
to $\mathcal{S}_{\langle ab\rangle}\theta^{\langle ab\rangle}$: Thus,
the flux of gravitational energy through the geoid is felt only through
shearing, which is precisely the effect of a linearly polarized gravitational
wave passing perpendicularly through the plane of an interferometric
detector.
\end{itemize}
We may add that this condition also renders comparison with RQFs easier
(since $\theta=0$ is one of the three RQF conditions (\ref{eq:RQFconditions})
as well). 

Moreover, imposing the GQF conditions (\ref{eq:GQFconditions}) turns
the Raychaudhuri equation (\ref{eq:Raychaudhuri}) into 
\begin{equation}
0=-\theta_{ab}\theta^{ba}-R_{ab}u^{a}u^{b}+\aleph\kappa+\boldsymbol{\mathcal{P}}^{2}+R_{acbd}n^{a}n^{b}u^{c}u^{d},\label{eq:RaychaudhuriGeoid}
\end{equation}
where we can use $\theta_{ab}\theta^{ba}=\theta_{\langle ab\rangle}\theta^{\langle ab\rangle}-\theta_{\left[ab\right]}\theta^{\left[ab\right]}$
for $\theta=0$. A simplified case of interest might be when there
is no rotation and the metric of $\mathscr{M}$ is a vacuum solution
of the Einstein equation; then, the above reduces to $0=-\theta_{\langle ab\rangle}\theta^{\langle ab\rangle}+\aleph\kappa+\boldsymbol{\mathcal{P}}^{2}+R_{acbd}n^{a}n^{b}u^{c}u^{d}$.

\section{Stationary Axisymmetric Solutions}
\label{sec:Solns}

Given a spacetime $\left(\mathscr{M},g\right)$ with a decomposition
perscribed by subsection \ref{sub:Spacetime-Decomposition}, GQFs
can in general be constructed by imposing the conditions given in the preceding section, that is to say, by solving (in
the adapted coordinates) the system of three first-order PDEs $0=\partial_{\mathfrak{i}}N+\dot{\hat{u}}_{\mathfrak{i}}$
and $0=\hat{\sigma}^{\mathfrak{ij}}\dot{\hat{\sigma}}_{\mathfrak{ij}}$.

If the spacetime is \textsl{stationary}, the problem reduces to solving
just two equations, $0=\partial_{\mathfrak{i}}N$. (Assuming $u^{a}$
is hypersurface orthogonal, $\mathscr{H}$ will be integrable into closed 2-dimensional hypersurfaces -- as
shown in Figure \ref{fig:QF}.)

If moreover the spacetime is \textsl{axisymmetric}, one of the two equations
$0=\partial_{\mathfrak{i}}N$ is automatically satisfied and thus
the problem is reduced to solving just one PDE: namely, requiring
that the partial of the lapse with respect to the polar angular coordinate
vanishes. (The partial with respect to the azimuthal coordinate automatically vanishes in such cases.)

A general method for constructing geoids in stationary axisymmetric
spacetimes is therefore as follows: If $g_{ab}$ is given in spherical
coordinates $\left(T,R,\Theta,\Phi\right)$, apply to it a coordinate
transformation
\begin{equation}
\begin{cases}
T & =t,\\
R & =f\left(r,\theta\right),\\
\Theta & =\theta,\\
\Phi & =\phi,
\end{cases}\label{eq:CoordTrans}
\end{equation}
and then solve $0=\partial_{\theta}N$ for the unknown function $f\left(r,\theta\right)$.
This will yield a one-parameter family (i.e. parametrized by the value
of $r$) of axisymmetric (i.e. constant in $\phi$) two-surfaces.

Moreover, after changing to $\left(t,r,\theta,\phi\right)$ coordinates,
we employ 
\begin{equation}
u^{a}=\left(\frac{1}{N}\right)\delta^{a}\,_{t},\quad n_{a}=\left(\frac{1}{\sqrt{g^{rr}}}\right)\delta^{r}\,_{a},\label{eq:unAdapted}
\end{equation}
to compute $\gamma_{ab}$, $\sigma_{ab}$, and all other related quantities
defined in Section \ref{sec:QF}.

For the rest of this section, we present GQF solutions corresponding to some stationary axisymmetric spacetimes of interest -- for Schwarzschild, excluding and including deformations, for slow rotation, and for Kerr.

\subsection{Schwarzschild Metric}

As a first application, let us consider the standard Schwarzschild
metric, 
\begin{equation}
{\rm d}s^{2}=-\left(1-\frac{2M}{R}\right){\rm d}T^{2}+\left(1-\frac{2M}{R}\right)^{-1}{\rm d}R^{2}+R^{2}\left({\rm d}\Theta^{2}+\sin^{2}\Theta{\rm d}\Phi^{2}\right).\label{eq:SchwMetric}
\end{equation}
Applying (\ref{eq:CoordTrans}), we get $N=\sqrt{1-2M/f\left(r,\theta\right)}$.
The GQF condition $0=\partial_{\theta}N$ then entails that $f$ has
no angular dependence, i.e. $f\left(r,\theta\right)=F\left(r\right)$
and so we simply have 

\begin{equation}
\boxed{R=F\left(r\right).}\label{eq:RsolnSchw}
\end{equation}The
degree of freedom in choosing this function can be regarded as a
general-relativistic version of the freedom we have in relabeling the family of geoids (which, in NG, is manifested via the more limited freedom available to fix boundary
conditions for the potential, i.e. to simply add to it an arbitrary constant).

With this, we compute the strain rate tensor $\theta_{ab}$ and the
quasilocal momentum $\mathcal{P}^{a}$, and find that they are both
zero. Since $R_{ab}=0$, the Raychaudhuri GQF equation (\ref{eq:RaychaudhuriGeoid})
reduces to $0=\aleph\kappa+R_{acbd}n^{a}n^{b}u^{c}u^{d}.$
Concordantly, we compute
\begin{equation}
\aleph=\frac{M}{F^{2}\left(r\right)\sqrt{1-2M/F\left(r\right)}},\quad\kappa=\frac{2\sqrt{1-2M/F\left(r\right)}}{F\left(r\right)},\quad R_{acbd}n^{a}n^{b}u^{c}u^{d}=-\frac{2M}{F^{3}\left(r\right)},\label{eq:RaychTermsSchw}
\end{equation}
thus verifying that the Raychaudhuri GQF equation is indeed satisfied in this case; what it is saying is that the normal acceleration ($\aleph$)
of observers on a geoidal surface with extrinsic curvature ($\kappa$),
which the observers would expect to cause them to diverge from each
other, is exactly balanced by their tidal acceleration (the $R_{acbd}n^{a}n^{b}u^{c}u^{d}$
term).

\subsection{Arbitrarily Deformed Schwarzschild Metric}

We now consider the Schwarzschild metric with arbitrary internal and
external deformations, i.e. with multipole moments of the source and
of the external field respectively. In prolate spheroidal coordinates,
it can be written as \cite{Breton}: 
\begin{equation}
{\rm d}s^{2}=-{\rm e}^{2\psi}{\rm d}T^{2}+M^{2}{\rm e}^{-2\psi}\left[{\rm e}^{2\gamma}\left(X^{2}-Y^{2}\right)\left(\frac{{\rm d}X^{2}}{X^{2}-1}+\frac{{\rm d}Y^{2}}{1-Y^{2}}\right)+\left(X^{2}-1\right)\left(1-Y^{2}\right){\rm d}\Phi^{2}\right],\label{eq:DeformedSchwMetric}
\end{equation}
where $X$ and $Y$ are defined\footnote{By equation (4) in \cite{Breton}: $X=\left(R_{+}+R_{-}\right)/2M$ and $Y=\left(R_{+}-R_{-}\right)/2M$,
with $R_{\pm}=\left[R^{2}+M^{2}\pm2MR\cos\Theta\right]^{1/2}$.} in terms of $R$ and $\Theta$, $\gamma$
is a complicated function\footnote{Given by equations (8) and (9)
in \cite{Breton}, which are long and we refrain from reproducing them here.} of $X$ and $Y$, and the function $\psi$ has the form 
\begin{equation}
\psi=\frac{1}{2}\ln\left(\frac{X-1}{X+1}\right)-\sum_{n=1}^{\infty}\left(\frac{A_{n}}{R^{n+1}}+B_{n}R^{n}\right){\rm P}_{n}\left(\frac{XY}{R}\right).\label{eq:DeformedSchwPsi}
\end{equation}
Here, the first term gives the undeformed Schwarzschild solution, the $A_{n}$
terms in the sum describe deformations of the source, and the $B_{n}$
terms describe deformations of the external field (with ${\rm P}_{n}$
denoting the $n$-th Legendre polynomial). To satisfy
certain regularity conditions \cite{Breton} we can discard all the
odd terms and retain only the even multipole moments; then applying (\ref{eq:CoordTrans}) we
get: 
\begin{equation}
N=\exp\left[\frac{1}{2}\ln\left(1-\frac{2M}{f\left(r,\theta\right)}\right)-\sum_{n=1}^{\infty}\left(\frac{A_{2n}}{f^{2n+1}\left(r,\theta\right)}+B_{2n}f^{2n}\left(r,\theta\right)\right){\rm P}_{2n}\left(\cos\theta\right)\right].\label{eq:DeformedSchwLapse}
\end{equation}
The GQF condition $0=\partial_{\theta}N$ is therefore solved implicitly
by $R=f\left(r,\theta\right)$ satisfying the algebraic equation 
\begin{equation}
\boxed{V_{{\rm R}}\left(r\right)=\frac{1}{2}\ln\left(1-\frac{2M}{R}\right)-\sum_{n=1}^{\infty}\left(\frac{A_{2n}}{R^{2n+1}}+B_{2n}R^{2n}\right){\rm P}_{2n}\left(\cos\theta\right),}\label{eq:DeformedSchwGQFsoln}
\end{equation}
where we have (suggestively) denoted $V_{{\rm R}}$ an arbitrary function
of $r$.

To see how this relates to the Newtonian concept of the geoid (``equipotential
surface''), we can Taylor expand (outside the Schwarzschild radius) $\ln\left(1-2M/R\right)=-\sum_{j=1}^{\infty}\left(2M/R\right)^{j}/j$, and restore units
to get 
\begin{equation}
V_{{\rm R}}=V_{{\rm N}}-\sum_{j=1}^{\infty}\left[\frac{2^{j}\left(GM\right)^{j+1}}{\left(j+1\right)c^{2j}}\right]\frac{1}{R^{j+1}},\label{eq:DeformedSchwGQFsolnSeries}
\end{equation}
where 
\begin{equation}
V_{{\rm N}}=-\frac{GM}{R}-\sum_{n=1}^{\infty}\left(\frac{A_{2n}}{R^{2n+1}}+B_{2n}R^{2n}\right){\rm P}_{2n}\left(\cos\theta\right)\label{eq:NewtonianPotential}
\end{equation}
is the usual (azimuthally-symmetric) solution to the Laplace equation in $\mathbf{R}^{3}$ i.e. simply the Newtonian potential ($-GM/R$
plus higher multipole corrections). Thus, $V_{{\rm R}}=V_{{\rm N}}+\mathcal{O}(1/c^{2})$
can be interpreted as being a general-relativistic version of the
``gravitational potential,'' with the sum (over $j$) on the RHS of (\ref{eq:DeformedSchwGQFsolnSeries})
supplying the relativistic corrections (beginning at $\mathcal{O}(1/c^{2})$).

\subsection{Slow-Rotation Metric}

Next we turn to including the effects of rotation. Before looking at the full Kerr metric, let us consider first its approximation
in the case of slow rotation, i.e. to first order in the angular momentum $J$. This is given by \cite{Adler}: 
\begin{equation}
{\rm d}s^{2}=-\left(1-\frac{2M}{R}\right){\rm d}T^{2}+\left(1-\frac{2M}{R}\right)^{-1}{\rm d}R^{2}+R^{2}\left({\rm d}\Theta^{2}+\sin^{2}\Theta{\rm d}\Phi^{2}\right)+\frac{4Ma}{R}\sin^{2}\Theta{\rm d}T{\rm d}\Phi,\label{eq:SlowRotationMetric}
\end{equation}
where $a=-J/M$. Into this we again insert (\ref{eq:CoordTrans}), and get $N=\sqrt{1-2M/f\left(r,\theta\right)}$.
Hence, just like in the (undeformed) Schwarzschild case, we simply
have 
\begin{equation}
\boxed{R=F\left(r\right).}\label{eq:RsolnSlowRotation}
\end{equation}Thus, effects
on the geoid due to rotation are only felt at $\mathcal{O}(a^{2})$.

\subsection{Kerr Metric}

We now consider the Kerr metric,
\begin{multline}
{\rm d}s^{2}=-\left(1-\frac{2MR}{R^{2}+a^{2}\cos^{2}\Theta}\right){\rm d}T^{2}+\left(\frac{R^{2}+a^{2}\cos^{2}\Theta}{R^{2}+a^{2}-2MR}\right){\rm d}R^{2}+\left(R^{2}+a^{2}\cos^{2}\Theta\right){\rm d}\Theta^{2}\\
+\left[\left(R^{2}+a^{2}\right)\sin^{2}\Theta+\frac{2MRa^{2}\sin^{4}\Theta}{R^{2}+a^{2}\cos^{2}\Theta}\right]{\rm d}\Phi^{2}+2\left(\frac{2MRa\sin^{2}\Theta}{R^{2}+a^{2}\cos^{2}\Theta}\right){\rm d}T{\rm d}\Phi.\label{eq:KerrMetric}
\end{multline}
Performing the coordinate transformation (\ref{eq:CoordTrans}), we
find
\begin{equation}
N=\sqrt{1-\frac{2Mf\left(r,\theta\right)}{f^{2}\left(r,\theta\right)+a^{2}\cos^{2}\theta}}.\label{eq:KerrLapse}
\end{equation}
The GQF condition $0=\partial_{\theta}N$ yields the following PDE
for $f$:
\begin{equation}
0=\left(f^{2}-a^{2}\cos^{2}\theta\right)\partial_{\theta}f-2a^{2}\left(\sin\theta\cos\theta\right)f.\label{eq:KerrGQFeqn}
\end{equation}
The general solution to this is $f_{\pm}\left(r,\theta\right)=\frac{1}{2}(F\left(r\right)\pm\sqrt{F^{2}\left(r\right)-4a^{2}\cos^{2}\theta}).$
To see which solution we should use, we expand each in powers of $a$:
we get $f_{-}\left(r,\theta\right)=\left[\left(\cos^{2}\theta\right)/F\left(r\right)\right]a^{2}+\mathcal{O}\left(a^{4}\right)$,
while $f_{+}\left(r,\theta\right)=F\left(r\right)-\left[\left(\cos^{2}\theta\right)/F\left(r\right)\right]a^{2}+\mathcal{O}\left(a^{4}\right)$.
But we know that in the slow-rotation limit, viz. the previous subsection,
we should just recover $R=f\left(r\right)$. Therefore, we can simply
discard the solution with the minus sign. We thus have: 
\begin{equation}
\boxed{R=\frac{1}{2}\left(F\left(r\right)+\sqrt{F^{2}\left(r\right)-4a^{2}\cos^{2}\theta}\right).}\label{eq:KerrGQFsoln}
\end{equation}
As $a\rightarrow0$, we recover the Schwarzschild GQF. Moreover, we
remark that (\ref{eq:KerrGQFsoln}) implies that Kerr GQFs must satisfy
$a^{2}\leq\frac{1}{4}F^{2}\left(r\right)$, i.e. they do not exist
for large enough angular momentum. Moreover, if $a^{2}<\frac{1}{4}F^{2}\left(r\right)$,
the geoid is smooth across the poles (i.e. $\partial_{\theta}f=0$
at $\theta=0,\pi$) but if $a^{2}=\frac{1}{4}F^{2}\left(r\right)$
(the maximally allowed value) then there will be a cusp there.

\begin{figure}
\noindent \begin{centering}
\includegraphics[scale=0.4]{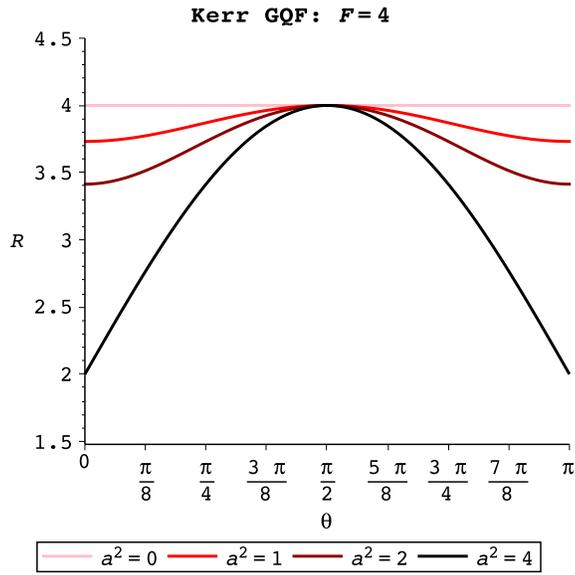} 
\par\end{centering}

\protect\protect\protect\caption{\label{fig:2d_plot_kerr_F=00003D00003D4}The radial function for the
Kerr GQF with $F\left(r\right)=r=4M$ in units of $M=1$, for different
values of the angular momentum parameter $a$. (Increasing colour
darkness corresponds to increasing $a$.)}
\end{figure}

\begin{figure}
\noindent \begin{centering}
\includegraphics[scale=0.4]{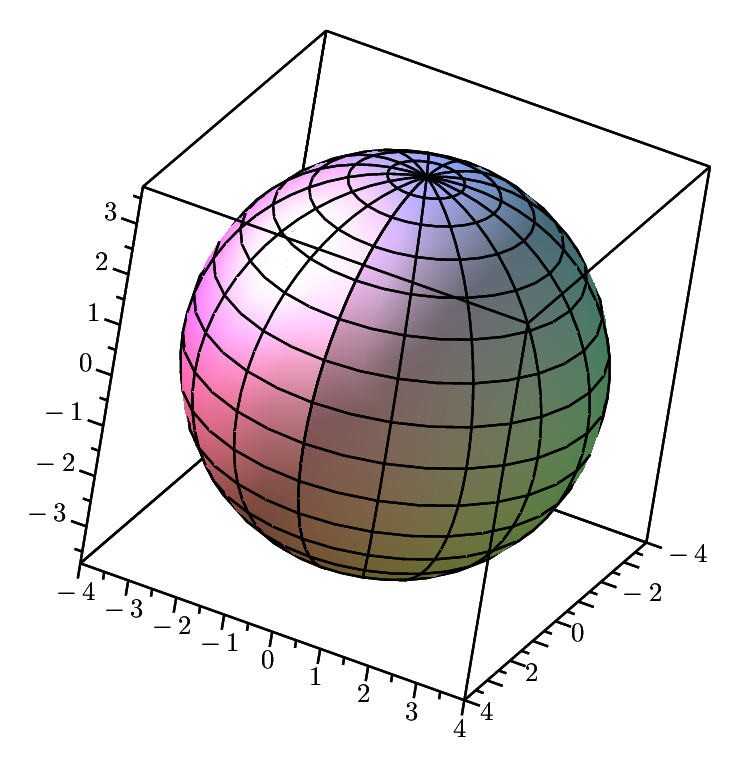}\includegraphics[scale=0.42]{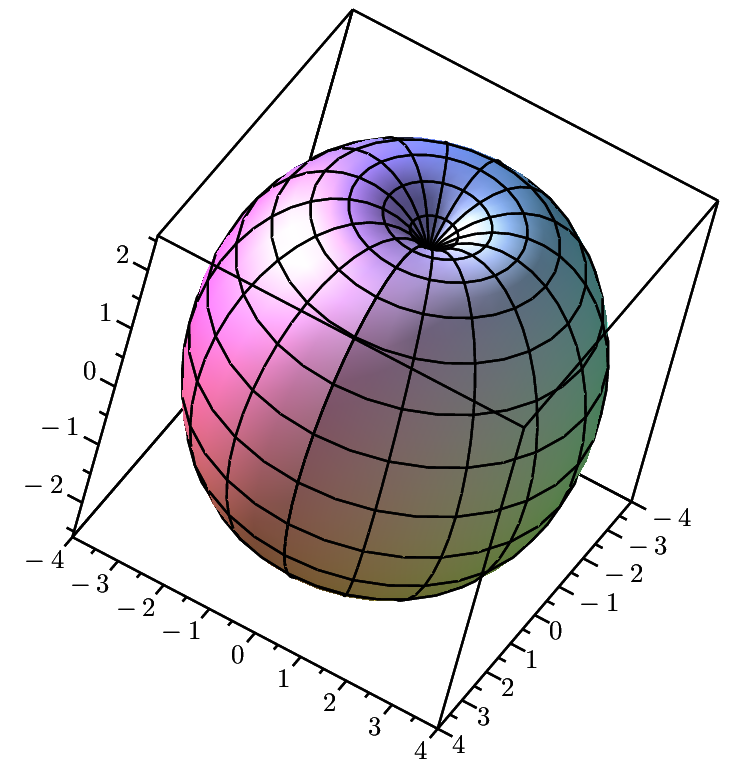} 
\par\end{centering}

\protect\protect\caption{\label{fig:3d_plot_kerr_F=00003D00003D4}The two-surface for the Kerr
GQF with $F\left(r\right)=r=4M$ in units of $M=1$, for angular momentum
parameter $a^{2}=1$ (left) and $a^{2}=4$ (right).}
\end{figure}

Setting $F\left(r\right)=r$, we plot (\ref{eq:KerrGQFsoln}) for
$r=4M$ in units of $M=1$ in Figure \ref{fig:2d_plot_kerr_F=00003D00003D4};
we also plot the resulting two-surfaces for different values of $a^{2}$ in
Figure \ref{fig:3d_plot_kerr_F=00003D00003D4}.

Furthermore, we analyze the Raychaudhuri GQF equation (\ref{eq:RaychaudhuriGeoid})
with the solution (\ref{eq:KerrGQFsoln}) in Appendix \ref{sec:B}; in particular, we compute each term in it exactly and check that it is satisfied.

~

~

~

~

\section{Geoids in General Relativity versus Newtonian Gravity}
\label{sec:NGvGR}

In this entire section, we work for simplicity in units of $M=1$.

For the purposes of quantifying and describing more concretely some
measurable discrepancies between a geoid according to GR and the corresponding
solution one would find in NG, we restrict our attention to the Schwarzschild
metric with only the leading order inner multipole, i.e. the inner
quadrupole moment, and set $B_{2}=0$, $A_{2n}=0=B_{2n}$ $\forall n\geq2$
(in which case (\ref{eq:DeformedSchwMetric}) is known as the \textsl{Erez-Rosen
metric}). Moreover, we treat the quadrupole moment $A_{2}:=Q$ as a
perturbation. Then, Taylor expanding in $Q$, (\ref{eq:DeformedSchwMetric})
can be written in the much more tractable form \cite{Frutos-Alfaro:2014usa}:
\begin{equation}
{\rm d}s^{2}=-{\rm e}^{-2\chi}\left(1-2/R\right){\rm d}T^{2}+{\rm e}^{2\chi}\left[\left(1-2/R\right)^{-1}{\rm d}R^{2}+R^{2}\left({\rm d}\Theta^{2}+\sin^{2}\Theta{\rm d}\Phi^{2}\right)\right],\label{eq:PerturbedERMetric}
\end{equation}
where 
\begin{equation}
\chi\left(R,\Theta\right)=\frac{Q}{R^{3}}{\rm P}_{2}\left(\cos\Theta\right).\label{eq:ERChi}
\end{equation}
This metric is valid up to $\mathcal{O}\left(Q\right)$.

We seek to compare geoid solutions
of (\ref{eq:PerturbedERMetric}) with geoids (``equipotential surfaces")
in NG also having a quadrupole moment $Q$. The latter
are just closed two-dimensional surfaces (topologically $\mathbf{S}^{2}$)
in $\mathbf{R}^{3}$, whose radius at any given polar angle (assuming
azimuthal symmetry) is determined by setting the solution to the Laplace
equation to a constant, i.e. $\mathscr{H}=\left\{ \left(R,\Theta,\Phi\right)\in\mathbf{R}^{3}|V_{{\rm N}}=-1/R-\left(Q/R^{3}\right){\rm P}_{2}\left(\cos\Theta\right)={\rm const.}\right\}$.
We will compute, in turn -- and then compare -- some geometric quantities
corresponding to the Newtonian and the Schwarzschild geoid with a
quadrupole perturbation.

\begin{enumerate}
\item \textbf{\textsl{NG}}: We use serif indices for (spherical) coordinate
components in $\mathbf{R}^{3}$, and (as before) Fraktur indices for
the geoid 2-surface $\mathscr{H}\simeq\mathbf{S}^{2}$. We begin with
the Euclidean 3-metric in spherical coordinates, 
\begin{equation}
{\rm d}s^{2}={\rm d}R^{2}+R^{2}\left({\rm d}\Theta^{2}+\sin^{2}\Theta{\rm d}\Phi^{2}\right),\label{eq:EuclideanSphericalMetric}
\end{equation}
and apply the coordinate transformation $R=f\left(r,\theta\right)$,
$\Theta=\theta$ and $\Phi=\phi$. Then, we compute the unit normal
vector to be $n_{\mathsf{a}}=(\left(f\partial_{r}f\right)/[f^{2}+\left(\partial_{\theta}f\right)^{2}]^{1/2},0,0),$
yielding the induced metric 
\begin{equation}
\sigma_{\mathsf{ab}}=\left(\begin{array}{ccc}
\left(\partial_{r}f\right)^{2}\left(\partial_{\theta}f\right)^{2}/\left[f^{2}+\left(\partial_{\theta}f\right)^{2}\right] & \left(\partial_{r}f\right)\left(\partial_{\theta}f\right) & 0\\
\left(\partial_{r}f\right)\left(\partial_{\theta}f\right) & f^{2}+\left(\partial_{\theta}f\right)^{2} & 0\\
0 & 0 & \left(\sin^{2}\theta\right)f^{2}
\end{array}\right).\label{eq:NGsigma_abQuadrupole}
\end{equation}
Here, $f$ is the function that is the solution to the (algebraic)
equation 
\begin{equation}
V_{{\rm N}}=-\frac{1}{f}-\frac{Q}{f^{3}}{\rm P}_{2}\left(\cos\theta\right).\label{eq:NGPotentialQuadrupole}
\end{equation}
Differentiating (\ref{eq:NGPotentialQuadrupole}) with respect to $\theta$
(with $V_{{\rm N}}$ fixed) yields an expression for $\partial_{\theta}f$
simply in terms of $f$. We can then substitute this, for example, into the resulting
2-metric on $\mathscr{H}$, 
\begin{equation}
\hat{\sigma}_{\mathfrak{ij}}=\left(\begin{array}{cc}
f^{2}+\left(\partial_{\theta}f\right)^{2} & 0\\
0 & \left(\sin^{2}\theta\right)f^{2}
\end{array}\right).\label{eq:NGsigma_ijQuadrupole}
\end{equation}
Moreover, we also compute $\kappa_{\mathsf{ab}}=\sigma_{\mathsf{ac}}\nabla^{\mathsf{c}}n_{\mathsf{b}}$
and projecting it onto $\mathscr{H}$ we get: 
\begin{equation}
\hat{\kappa}_{\mathfrak{ij}}=\frac{1}{\sqrt{f^{2}+\left(\partial_{\theta}f\right)^{2}}}\left(\begin{array}{cc}
f^{2}+2\left(\partial_{\theta}f\right)^{2}-f\partial_{\theta}^{2}f & 0\\
0 & \left(\sin\theta\right)\left[\left(\sin\theta\right)f-\left(\cos\theta\right)\partial_{\theta}f\right]f
\end{array}\right),\label{eq:NGkappa_ijQuadrupole}
\end{equation}
where we can express $\partial_{\theta}^{2}f$ by differentiating $\partial_{\theta}f$
again, and substituting the expression for it back in -- so that, once again, everything is purely in terms of $f$ (the solution to (\ref{eq:NGPotentialQuadrupole})). Finally, the
2-dimensional Ricci scalar (associated with $\hat{\sigma}_{\mathfrak{ij}}$)
is\footnote{N.B.: The second equality in (\ref{eq:NGRicciQuadrupole}) is simply the ``theorema egregium'' of
Gauss in 2 dimensions, which we see indeed holds.}
\begin{equation}
\mathfrak{R}=2\frac{\left[f^{2}+2\left(\partial_{\theta}f\right)^{2}-f\partial_{\theta}^{2}f\right]\left[\left(\sin\theta\right)f-\left(\cos\theta\right)\partial_{\theta}f\right]}{\left[f^{2}+\left(\partial_{\theta}f\right)^{2}\right]^{2}\left(\sin\theta\right)f}=2\frac{\det\left(\hat{\kappa}_{\mathfrak{ij}}\right)}{\det\left(\hat{\sigma}_{\mathfrak{ij}}\right)}.\label{eq:NGRicciQuadrupole}
\end{equation}

\item \textbf{\textsl{GR}}: Applying the coordinate transformation (\ref{eq:CoordTrans})
to (\ref{eq:PerturbedERMetric}) we compute the induced 2-metric on
$\mathscr{H}$ to be 
\begin{equation}
\hat{\sigma}_{\mathfrak{ij}}={\rm e}^{2\chi\left(f,\theta\right)}\left(\begin{array}{cc}
\left[f^{2}+\left(\partial_{\theta}f\right)^{2}-2Mf\right]f/\left(f-2M\right) & 0\\
0 & \left(\sin^{2}\theta\right)f^{2}
\end{array}\right).\label{eq:ERsigma_ij}
\end{equation}
Here, $f$ is the function that solves the (algebraic) equation 
\begin{equation}
V_{{\rm R}}=\frac{1}{2}\ln\left(1-\frac{2}{f}\right)-\frac{Q}{f^{3}}{\rm P}_{2}\left(\cos\theta\right).\label{eq:GRPotentialQuadrupole}
\end{equation}
with $V_{{\rm R}}$ constant. From (\ref{eq:GRPotentialQuadrupole})
we get (just like in the NG case) an expression for $\partial_{\theta}f$ in terms of $f$ by
taking a partial with respect to $\theta$, which (along with the
resulting expression for $\partial_{\theta}^{2}f$ from differentiating
this again) we can use to express $\hat{\sigma}_{\mathfrak{ij}}$
as well as all other geometric quantities such as $\hat{\kappa}_{\mathfrak{ij}}$
and $\mathfrak{R}$ purely in terms of $f$, which in turn is just
the solution to (\ref{eq:GRPotentialQuadrupole}). 
\end{enumerate}
For the purposes of finding differences between what NG and GR geoids, it is instructive to compare the Ricci scalar predicted by the
two theories (with other quantities held fixed). It is also instructive
to compare not the extrinsic curvature itself, but the proper components
of the extrinsic curvature, that is the non-vanishing components of
$\kappa_{IJ}=e_{I}\,^{a}e_{J}\,^{b}\kappa_{ab}$, where $\left\{ e_{I}\,^{a}\right\} $
is a unit dyad tangent to $\mathscr{H}$. We take $e_{1}\,^{a}=\left(1/\sqrt{\hat{\sigma}_{\theta\theta}}\right)\left(\partial_{\theta}\right)^{a}$
and $e_{2}\,^{a}=\left(1/\sqrt{\hat{\sigma}_{\phi\phi}}\right)\left(\partial_{\phi}\right)^{a}$
(so that $\bm{e}_{1}\cdot\bm{e}_{1}=1=\bm{e}_{2}\cdot\bm{e}_{2}$),
and thus the proper extrinsic curvature along $\partial_{\theta}$
is $k_{1}:=\kappa_{11}=\hat{\kappa}_{\theta\theta}/\hat{\sigma}_{\theta\theta}$
and, similarly, along $\partial_{\phi}$ it is $k_{2}:=\kappa_{22}=\hat{\kappa}_{\phi\phi}/\hat{\sigma}_{\phi\phi}$.
(The off-diagonal component is zero.)

Following the above procedure, we have found exact expressions (as
functions of $\theta$, in terms of $Q$ and $f$ or, implicitly, the potential
value) for all the relevant geometric quantities ($\mathfrak{R}$,
$k_{1}$ and $k_{2}$). They are very cumbersome
to write down in full, so instead we offer here their series expressions
up to linear order in $Q$ (for NG as well as GR). For notational convenience, we henceforth denote
$\mathbb{P}\left(\theta\right):={\rm P}_{2}\left(\cos\theta\right)=\frac{1}{2}\left(3\cos^{2}\theta-1\right)$.
\begin{enumerate}
\item \textbf{\textsl{NG}}: We get
\begin{alignat}{1}
\mathfrak{R} & =\frac{2}{f^{2}}+\left\{ \frac{12\mathbb{P}\left(\theta\right)}{f^{4}}\right\} Q+\ldots\label{eq:QseriesNG_R}\\
 & =2V_{{\rm N}}^{2}+\left\{ 8V_{{\rm N}}^{4}\mathbb{P}\left(\theta\right)\right\} Q+\ldots\label{eq:eq:QseriesNG_R_2}\\
k_{1} & =\frac{1}{f}+\left\{ \frac{3\left(2\cos^{2}\theta-1\right)}{f^{3}}\right\} Q+\ldots\label{eq:QseriesNG_k1}\\
 & =-V_{{\rm N}}+\left\{ -\frac{1}{2}V_{{\rm N}}^{3}\left(9\cos^{2}\theta-5\right)\right\} Q+\ldots\label{eq:QseriesNG_k1_2}\\
k_{2} & =\frac{1}{f}+\left\{ \frac{3\cos^{2}\theta}{f^{3}}\right\} Q+\ldots\label{eq:QseriesNG_k2}\\
 & =-V_{{\rm N}}+\left\{ -\frac{1}{2}V_{{\rm N}}^{3}\left(3\cos^{2}\theta+1\right)\right\} Q+\ldots\label{eq:QseriesNG_k2_2}
\end{alignat}
where $\ldots$ denotes $\mathcal{O}\left(Q^{2}\right)$ terms and,
to get the second lines for each quantity, we have used 
\begin{equation}
f=-\frac{1}{V_{{\rm N}}}+\left\{ -V_{{\rm N}}\mathbb{P}\left(\theta\right)\right\} Q+\ldots\label{eq:QseriesNG_f}
\end{equation}
This is obtained by inserting a series ansatz $f=c_{0}+c_{1}Q+\ldots$
into (\ref{eq:NGPotentialQuadrupole}), and solving for the coefficients
to satisfy the equation order-by-order in $Q$.
\item \textbf{\textsl{GR}}: We get
\begin{alignat}{1}
\mathfrak{R} & =\frac{2}{f^{2}}+\left\{ \frac{4\left(3f-4\right)\mathbb{P}\left(\theta\right)}{f^{5}}\right\} Q+\ldots\label{eq:QseriesGR_R}\\
 & =\frac{1}{2}\left(1-{\rm e}^{2V_{{\rm R}}}\right)^{2}+\left\{ \frac{1}{4}\left(1-{\rm e}^{4V_{{\rm R}}}\right)\left(1-{\rm e}^{2V_{{\rm R}}}\right)^{3}\mathbb{P}\left(\theta\right)\right\} Q+\ldots\label{eq:QseriesGR_R_2}\\
k_{1} & =\frac{\sqrt{1-2/f}}{f}+\left\{ \frac{\sqrt{1-2/f}\left[6\left(f-1\right)\cos^{2}\theta-3f+2\right]}{f^{4}}\right\} Q+\ldots\label{eq:QseriesGR_k1}\\
 & =\frac{1}{2}{\rm e}^{V_{{\rm R}}}\left(1-{\rm e}^{2V_{{\rm R}}}\right)+\left\{ \frac{1}{32}{\rm e}^{V_{{\rm R}}}\left(1-{\rm e}^{2V_{{\rm R}}}\right)^{3}\left(-{\rm e}^{2V_{{\rm R}}}+15\cos^{2}\theta-9\right)\right\} Q+\ldots\label{eq:QseriesGR_k1_2}\\
k_{2} & =\frac{\sqrt{1-2/f}}{f}+\left\{ \frac{\sqrt{1-2/f}\left[3\left(f-2\right)\cos^{2}\theta+2\right]}{f^{4}}\right\} Q+\ldots\label{eq:QseriesGR_k2}\\
 & =\frac{1}{2}{\rm e}^{V_{{\rm R}}}\left(1-{\rm e}^{2V_{{\rm R}}}\right)+\left\{ \frac{1}{32}{\rm e}^{V_{{\rm R}}}\left(1-{\rm e}^{2V_{{\rm R}}}\right)^{3}\left(2{\rm e}^{2V_{{\rm R}}}\mathbb{P}\left(\theta\right)+3\cos^{2}\theta+3\right)\right\} Q+\ldots\label{eq:QseriesGR_k2_2}
\end{alignat}
 where 
\begin{equation}
f=\frac{2}{1-{\rm e}^{2V_{{\rm R}}}}+\left\{ \frac{1}{2}{\rm e}^{2V_{{\rm R}}}\left(1-{\rm e}^{2V_{{\rm R}}}\right)\mathbb{P}\left(\theta\right)\right\} Q+\ldots\label{eq:QseriesGR_f}
\end{equation}
is here obtained from (\ref{eq:GRPotentialQuadrupole}) via the same
procedure as in the NG case.
\end{enumerate}
As a consistency check, notice that GR should recover NG when $V_{{\rm N}}\approx V_{{\rm R}}$
is small. Indeed, expanding the GR results (\ref{eq:QseriesGR_R_2}),
(\ref{eq:QseriesGR_k1_2}) and (\ref{eq:QseriesGR_k2_2}) further
in $V_{{\rm R}}$, we find $\mathfrak{R}=2V_{{\rm R}}^{2}+\mathcal{O}\left(Q^{2},V_{{\rm R}}^{3}\right)$,
$k_{1}=-V_{{\rm R}}+\mathcal{O}\left(Q^{2},V_{{\rm R}}^{2}\right)$
and $k_{2}=-V_{{\rm R}}+\mathcal{O}\left(Q^{2},V_{{\rm R}}^{2}\right)$,
thus indeed recovering the NG results (\ref{eq:eq:QseriesNG_R_2}), (\ref{eq:QseriesNG_k1_2}) and (\ref{eq:QseriesNG_k2_2}) respectively.

Now, to actually see how they compare, we plot the (full, exact) expressions
for each. To do this, we must first fix some quantity describing the
geoid ``size'' in all cases. We choose to fix the proper distance
$\Delta=\int{\rm d}s=\int{\rm d}\theta\sqrt{\hat{\sigma}_{\theta\theta}}$
travelled along a line of longitude of the geoid, from the North Pole
($\theta=0$) to the South Pole ($\theta=\pi$). In particular, in
all plots, we display results for a choice of $\Delta=\int_{0}^{\pi}{\rm d}\theta\sqrt{\hat{\sigma}_{\theta\theta}}=10$.
This is equivalent to a fixing of the potential values ($V_{{\rm N}}$
and $V_{{\rm R}}$ respecively), and we determine precisely which
potential values achieve a pole-to-pole proper distance of $10$ as
follows: 
\begin{itemize}
\item first, select the desired value of $Q$; 
\item then, perform a numerical integration of $\int_{0}^{\pi}{\rm d}\theta\sqrt{\hat{\sigma}_{\theta\theta}}$
via Simpson's rule (for which we used $100$ partitions), and retain
the result symbolically (that is, as sum that is a function of the
potential value); 
\item finally, apply Newton's method (for which we used $20$ iterations)
to find the potential value that is the root of the result of the previous step minus
$10$. For the initial guess, use the potential value corresponding to
the closest value of $Q$ previously obtained. 
\end{itemize}
In this way, we can plot everything for a pole-to-pole distance of
$10$.

Before displaying the results, we investigate what we should expect
for $Q=0$.
\begin{enumerate}
\item \textbf{\textsl{NG}}: (\ref{eq:NGPotentialQuadrupole}) implies $f=-1/V_{{\rm N}}$
for $Q=0$. Thus, for any constant pole-to-pole proper distance $\Delta=\int{\rm d}s$, we have
\begin{equation}
\Delta=\int_{0}^{\pi}{\rm d}\theta\sqrt{\hat{\sigma}_{\theta\theta}}=\int_{0}^{\pi}{\rm d}\theta f=-\frac{\pi}{V_{{\rm N}}}.\label{eq:pNG_Q=00003D0}
\end{equation}

\item \textbf{\textsl{GR}}: (\ref{eq:GRPotentialQuadrupole}) implies $f=2/\left(1-{\rm e}^{2V_{{\rm R}}}\right)$
for $Q=0$. So to get the same pole-to-pole proper distance $\Delta=\int{\rm d}s$,
we need
\begin{equation}
\Delta=\int_{0}^{\pi}{\rm d}\theta\sqrt{\hat{\sigma}_{\theta\theta}}=\int_{0}^{\pi}{\rm d}\theta f=\frac{2\pi}{1-{\rm e}^{2V_{{\rm R}}}}.\label{eq:pGR_Q=00003D0}
\end{equation}

\end{enumerate}
Setting (\ref{eq:pNG_Q=00003D0}) equal to (\ref{eq:pGR_Q=00003D0}),
we get $-\pi/\Delta=V_{{\rm N}}=-\frac{1}{2}\left(1-{\rm e}^{2V_{{\rm R}}}\right)$.
Inserting these into (\ref{eq:eq:QseriesNG_R_2})-(\ref{eq:QseriesGR_k2_2}),
we see that the Ricci scalar in NG and GR coincides
for $Q=0$ (as expected), and is given by
\begin{equation}
\mathfrak{R}=2V_{{\rm N}}^{2}=\frac{1}{2}\left(1-{\rm e}^{2V_{{\rm R}}}\right)^{2}=\frac{2\pi^{2}}{\Delta^{2}}.\label{eq:R_GR_Q=00003D0}
\end{equation}
However, the extrinsic geometries are different. In NG, we have for
$Q=0$,
\begin{equation}
k_{1}=k_{2}=-V_{{\rm N}}=\frac{1}{2}\left(1-{\rm e}^{2V_{{\rm R}}}\right)=\frac{\pi}{\Delta}.\label{eq:k_NG_Q=00003D0}
\end{equation}
Meanwhile, in GR,
\begin{equation}
k_{1}=k_{2}=\frac{1}{2}{\rm e}^{V_{{\rm R}}}\left(1-{\rm e}^{2V_{{\rm R}}}\right)=-\sqrt{1+2V_{{\rm N}}}V_{{\rm N}}=\frac{\pi}{\Delta}\sqrt{1-\frac{2\pi}{\Delta}}=\frac{\pi}{\Delta}-\frac{\pi^{2}}{\Delta^{2}}+\mathcal{O}\left(\frac{1}{\Delta^{3}}\right)\label{eq:k_GR_Q=00003D0}
\end{equation}
where in the last equality we have expanded in powers of $1/\Delta$.
(Hence, for large $\Delta$, the above recovers the NG result.) Thus
we might expect to be able to match the Ricci scalar (intrinsic geometry)
of a geoid in NG with GR, but not necessarily the extrinsic curvature
-- which, we see, is always lower in GR even with vanishing multipole
deformations.

\begin{figure}
\noindent \begin{centering}
\includegraphics[scale=0.4]{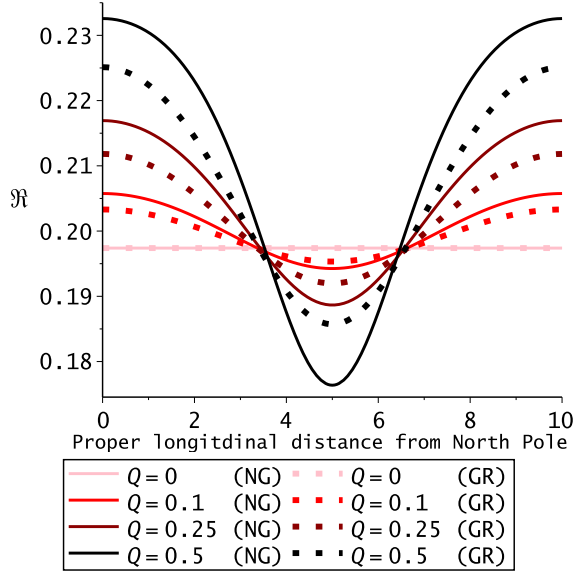} 
\par\end{centering}

\protect\protect\protect\protect\caption{\label{fig:2d_plot_NGvGR_Ricci_scalar_1}The induced 2-dimensional
Ricci scalar $\mathfrak{R}$ of the geoid 2-surface as a function
of the proper longitudinal distance travelled along it starting at
the North Pole. All plots are for a fixed pole-to-pole proper distance
of $10$, in both NG (solid) and GR (dotted), for different values
of the quadrupole moment $Q$ (increasing darkness corresponds to
increasing $Q$).}
\end{figure}

In Figure \ref{fig:2d_plot_NGvGR_Ricci_scalar_1}, we plot (parametrically
in $\theta$, for $0\leq\theta\leq\pi$) the induced 2-dimensional
Ricci scalar $\mathfrak{R}\left(\theta\right)$ of the geoid 2-surface
as a function of the proper longitudinal distance travelled along
the geoid starting at the North Pole $\int_{0}^{\theta}{\rm d}s$ (the integration again being performed numerically via Simpson's rule with $100$ partitions),
for a fixed pole-to-pole proper distance of $10$. This shows that
given the \textsl{same} geoid ``shape'' (measurement of the Ricci
scalar along the geoid), a Newtonian observer would deduce \textsl{different}
quadrupole -- and possibly higher multipole -- moments than a relativistic
observer would.

In particular, using the geoid map of, say, the Earth to infer its
quadrupole moment via the Newtonian theory would actually yield according
to this a value that is \textsl{lower} than the true one, i.e. the
value one would infer from a relativistic treatment of the (same)
geoid. The correction to $Q$ thus obtained may be important, for
example, in geophysical studies of the Earth's internal composition:
a higher $Q$ than previously assumed may indicate that the Earth
is in fact less well consolidated.

\begin{figure}
\noindent \begin{centering}
\includegraphics[scale=0.4]{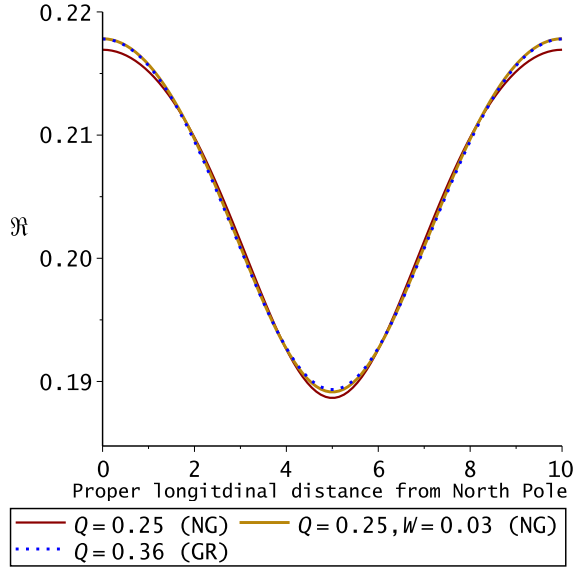} 
\par\end{centering}

\protect\protect\protect\protect\caption{\label{fig:2d_plot_NGvGR_Ricci_scalar_2}The induced 2-dimensional
Ricci scalar $\mathfrak{R}$ of the geoid 2-surface as a function
of the proper longitudinal distance travelled along it starting at
the North Pole, for a fixed pole-to-pole proper distance of $10$.
We display the quadrupole-only plot in NG for $Q=0.25$ (red solid),
the result of adding to this an octupole of $W=0.03$ (solid yellow),
and the GR plot for $Q=0.36$ (dotted blue). We see that the latter
two curves lie nearly on top of each other.}
\end{figure}

Moreover, we find that we obtain a nearly exact agreement between,
say, the $Q=0.25$ geoid in NG and the $Q=0.36$ geoid in GR by adding
to the former an octupole moment of $W=0.03$; this is achieved by
repeating the Newtonian analysis now with $V_{{\rm N}}=-1/f-\left(Q/f^{3}\right){\rm P}_{2}\left(\cos\theta\right)-\left(W/f^{5}\right){\rm P}_{4}\left(\cos\theta\right)$.
See the yellow and blue curves in Figure \ref{fig:2d_plot_NGvGR_Ricci_scalar_2}.
Thus the relativistic correction to the geoid is a genuinely nontrivial
effect, in that an observer may otherwise infer (different) \textsl{higher}
multipole contributions to resolve a given shape. Or, put differently,
what appear to be higher order multipole effects in NG are, actually,
simply a disguise for quadrupole corrections in GR.

\begin{figure}
\noindent \begin{centering}
\includegraphics[scale=0.4]{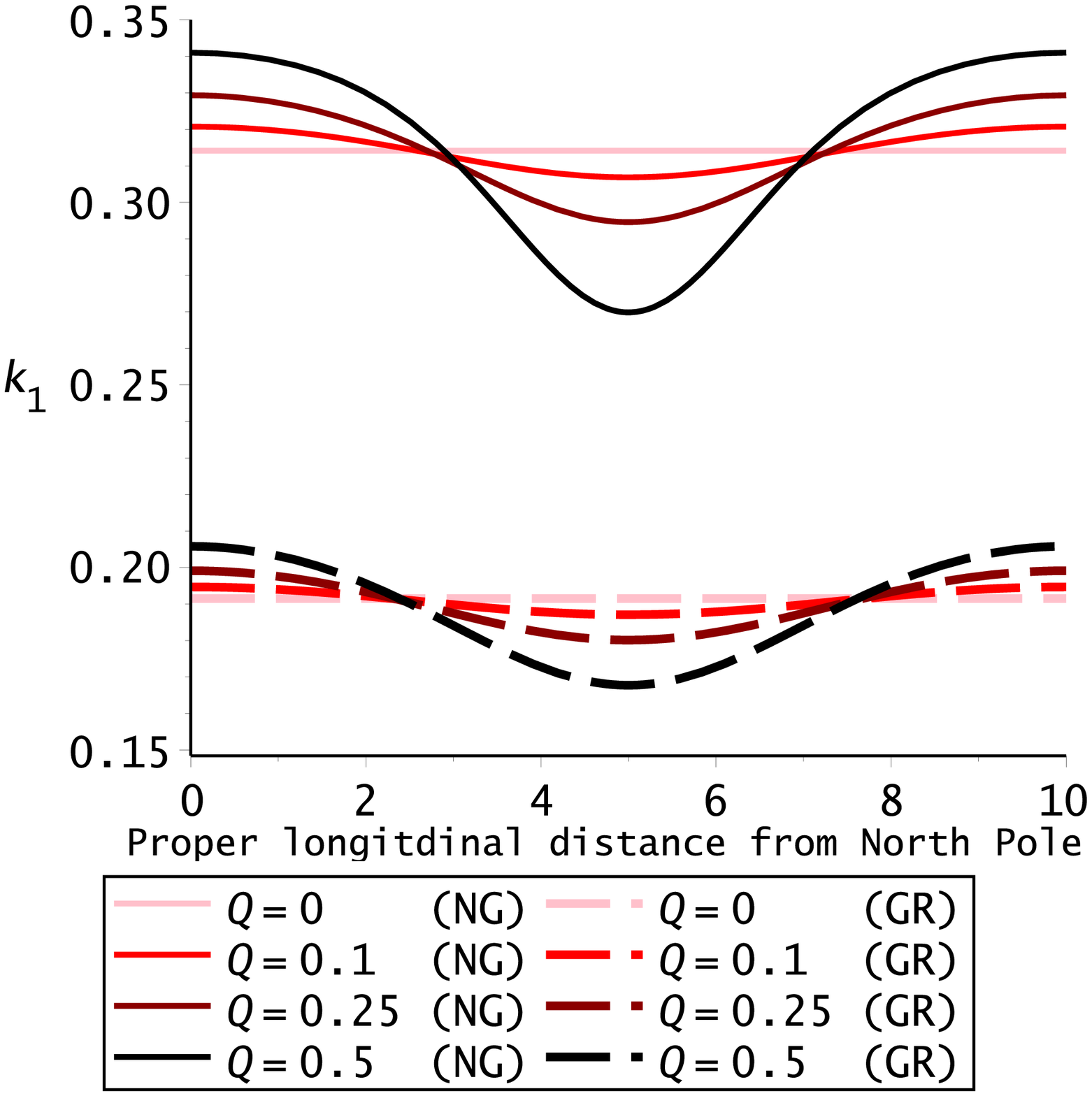} \includegraphics[scale=0.4]{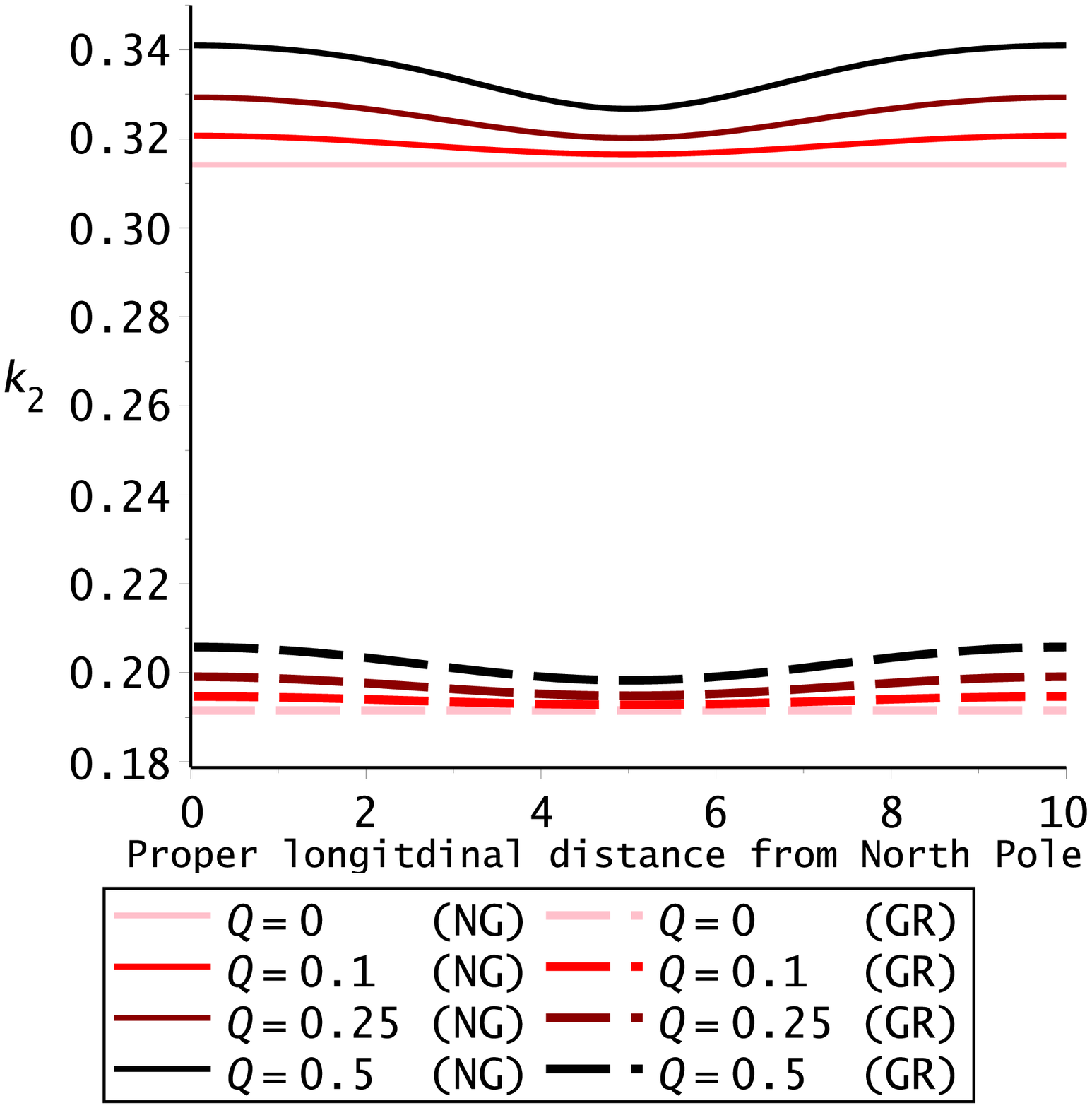} 
\par\end{centering}

\protect\protect\protect\protect\caption{\label{fig:2d_plot_NGvGR_k1_k2}The induced 2-dimensional
proper extrinsic curvatures $k_{1}$ and $k_{2}$ of the geoid 2-surface
as a function of the proper longitudinal distance travelled along
it starting at the North Pole. All plots are for a fixed pole-to-pole
proper distance of $10$, in both NG (solid) and GR (dotted), for
different values of the quadrupole moment $Q$ (increasing darkness
corresponds to increasing $Q$).}
\end{figure}

Finally, we plot $k_{1}$ and $k_{2}$ in Figure \ref{fig:2d_plot_NGvGR_k1_k2}. As expected, they
do not coincide for $Q=0$ and they cannot be made to agree (as was
the case with $\mathfrak{R}$) simply by the addition of multipole
effects.

\section{Conclusions}
\label{sec:Concl}

We have defined the notion of a geoid in GR, and have shown it to
fit naturally within a quasilocal approach for describing general extended systems in curved spacetime. Conservation laws can
be formulated in terms of them which may, as an obvious first application, prove easily amenable to computing
energy changes due to gravitational waves. Indeed, we have seen that the energy flux across a geoid is (in the absence of matter) simply given by shear stress times shear strain, which describes precisely the familiar effect of
a linearly polarized gravitational wave perpendicular to the plane of a detector. A detailed
inquiry is left to future work. 

Moreover, we
have constructed explicit solutions for geoids in some spacetimes,
and have compared a simple case among these -- namely, the Schwarzschild
solution containing a quadrupole perturbation -- with the NG geoid
containing the same. We have found that GR predicts nontrivial quadrupole corrections that, additionally to the actual value of the quadrupole moment itself, a Newtonian
treatment would mistake for a \textsl{higher-order} multipole effect. Such corrections will prove useful in applications of the geoid aiming to turn from a Newtonian to a fully general-relativistic approach -- as might well soon be the case with GPS.

Future work may include extending our analysis by including (possibly non-perturbatively) higher
multipoles in both the NG and GR geoid being compared, which (relative
to the perturbative quadrupole-only problem) is computationally much more involved. It may furthermore prove valuable to undertake the computation of other geometric quantities (aside from the Ricci scalar and extrinsic curvature) that are actually used in measuring the geoid map of the Earth, such as geodesic deviation. Solving the Kerr geoid with a quadrupole perturbation \cite{Frutos-Alfaro:2014usa} (and then possibly with arbitrary multipoles \cite{Breton}) would be another problem of immediate interest. Furthermore, it would also be valuable to carry out a post-Newtonian analysis of the geoid in the context of GQFs (analogously to that already done for RQFs \cite{McGrath:2013pea}) to augment/recast in a quasilocal setting the recent work \cite{Kopeikin:2015} in this direction by some of the same authors as \cite{Kopeikin:2014afa}.

\section*{Acknowledgements}

This work was supported by the Natural Sciences and Engineering Research
Council of Canada and the Ontario Ministry of Training, Colleges and
Universities. We thank Don Page for initially suggesting the GQF conditions
(\ref{eq:GQFconditions}) out of which this work has resulted. M.O.
would like to thank Dennis Phillip, Jason Pye and Robert H. Jonsson
for helpful discussions.

\renewcommand{\thesection}{A}
\section{The General Raychaudhuri Equation}\label{sec:A}

Here we show the main steps in the computation of (\ref{eq:RaychaudhuriGeoid}).
By the definition of the expansion tensor (\ref{eq:theta}) and the
Leibnitz rule,
\begin{align}
\nabla_{\boldsymbol{u}}\theta_{ab} & =u^{c}\nabla_{c}\left(\sigma_{ae}\sigma_{bf}\nabla^{e}u^{f}\right)\\
 & =u^{c}\left[\left(\nabla_{c}\sigma_{ae}\right)\sigma_{bf}+\sigma_{ae}\left(\nabla_{c}\sigma_{bf}\right)\right]+u^{c}\sigma_{ae}\sigma_{bf}\nabla_{c}\nabla^{e}u^{f}.
\end{align}
We insert $\sigma_{ab}=g_{ab}-n_{a}n_{b}+u_{a}u_{b}$ into the terms
in the square brackets, expand them out using the Leibnitz rule again,
and then multiply both sides of the equation with $g^{ab}$. Many
of the resulting terms cancel owing to the fact that $\sigma_{ab}n^{b}=0=\sigma_{ab}u^{b}$,
yielding
\begin{equation}
\frac{{\rm d}\theta}{{\rm d}\tau}=\sigma_{ab}\left(\nabla_{\boldsymbol{u}}u^{a}\right)\nabla_{\boldsymbol{u}}u^{b}-\sigma_{ab}\left(\nabla_{\boldsymbol{u}}n^{a}\right)\left(\nabla_{\boldsymbol{n}}u^{b}+n_{c}\nabla^{b}u^{c}\right)+g^{ab}u^{c}\sigma_{ae}\sigma_{bf}\nabla_{c}\nabla^{e}u^{f}.\label{eq:AUnsimplified}
\end{equation}
The first term on the RHS is, by definition, 
\begin{equation}
\sigma_{ab}\left(\nabla_{\boldsymbol{u}}u^{a}\right)\nabla_{\boldsymbol{u}}u^{b}=\boldsymbol{\alpha}\cdot\boldsymbol{a}=\boldsymbol{\alpha}\cdot\boldsymbol{\alpha}.\label{eq:AAlphaSquaredTerm}
\end{equation}
The second term can be rewritten, using once more the definition of
$\sigma_{ab}$ and orthogonality properties of $\boldsymbol{n}$ and
$\boldsymbol{u}$, in terms of the normal component of the acceleration
and the extrinsic curvature of $\mathscr{B}$ as:
\begin{equation}
-\sigma_{ab}\left(\nabla_{\boldsymbol{u}}n^{a}\right)\left(\nabla_{\boldsymbol{n}}u^{b}+n_{c}\nabla^{b}u^{c}\right)=-\left(\nabla_{\boldsymbol{u}}n_{b}\right)\nabla_{\boldsymbol{n}}u^{b}+\aleph^{2}+u^{a}u_{c}\Theta_{ab}\Theta^{bc}.\label{eq:AThetaabTerm}
\end{equation}
Finally, the third term can be expressed, first by applying the definition
of the Riemann tensor and the Leibnitz rule,
\begin{equation}
g^{ab}u^{c}\sigma_{a}\,^{e}\sigma_{b}\,^{f}\nabla_{c}\nabla_{e}u_{f}=g^{ab}\sigma_{a}\,^{e}\sigma_{bf}\left[\nabla_{e}\nabla_{\boldsymbol{u}}u^{f}-\left(\nabla_{e}u^{c}\right)\nabla_{c}u^{f}-R^{f}\,_{dec}u^{d}u^{c}\right],
\end{equation}
and then by using
\begin{equation}
\theta_{ab}\theta^{ba}=\sigma^{be}\sigma_{bf}\left(\nabla_{e}u^{c}\right)\nabla_{c}u^{f}-\sigma^{be}\sigma_{bf}\left[\left(\nabla_{\boldsymbol{n}}u^{f}\right)n^{d}\nabla_{e}u_{d}-\left(\nabla_{\boldsymbol{u}}u^{f}\right)u^{d}\nabla_{e}u_{d}\right],
\end{equation}
which follows from the definition of the expansion tensor (\ref{eq:theta}),
as
\begin{equation}
g^{ab}u^{c}\sigma_{ae}\sigma_{bf}\nabla_{c}\nabla^{e}u^{f}=-\theta_{ab}\theta^{ba}+\sigma^{ae}\sigma_{eb}\nabla_{a}a^{b}-\sigma^{ae}\sigma_{eb}\left(\nabla_{\boldsymbol{n}}u^{b}\right)n^{c}\nabla_{a}u_{c}-\sigma^{ae}\sigma_{eb}R^{b}\,_{dac}u^{d}u^{c}.\label{eq:AC}
\end{equation}
Now, using the Leibnitz rule, the definition of $\sigma_{ab}$, and
orthogonality properties of $\boldsymbol{n}$ and $\boldsymbol{u}$,
we can write the second term on the RHS of (\ref{eq:AC}) in terms
of the normal component of the acceleration and the trace of the extrinsic
curvature of $\mathscr{B}$ as: 
\begin{equation}
\sigma^{ae}\sigma_{eb}\nabla_{a}a^{b}=\sigma^{ae}\left(\nabla_{a}\alpha_{e}-a^{b}\nabla_{a}\sigma_{eb}\right)=\boldsymbol{\mathfrak{D}}\cdot\boldsymbol{\alpha}+\aleph\kappa=\boldsymbol{\mathfrak{D}}\cdot\boldsymbol{\alpha}+\aleph\left(\Theta-\aleph\right),\label{eq:AC1}
\end{equation}
where $\sigma^{ab}\nabla_{a}=\mathfrak{D}^{b}$, and in the last equality
we have used the useful fact (which can be shown from the definition
of $\sigma_{ab}$ and orthogonality properties) that $\Theta=\kappa+\aleph$.
The second term on the RHS of (\ref{eq:AC}), using analogous manipulations,
can be written as
\begin{equation}
-\sigma^{ae}\sigma_{eb}\left(\nabla_{\boldsymbol{n}}u^{b}\right)n^{c}\nabla_{a}u_{c}=\Theta_{ab}u^{a}\nabla_{\boldsymbol{n}}u^{b}=\left(\nabla_{\boldsymbol{u}}n_{b}\right)\nabla_{\boldsymbol{n}}u^{b}.\label{eq:AC2}
\end{equation}
Using the definition of $\sigma_{ab}$ and symmetry properties of
the Riemann tensor, the last term on the RHS of (\ref{eq:AC}) is
\begin{equation}
-\sigma^{ae}\sigma_{eb}R^{b}\,_{dac}u^{d}u^{c}=-R_{ab}u^{a}u^{b}+R_{acbd}n^{a}n^{b}u^{c}u^{d}.\label{eq:AC3}
\end{equation}
Inserting (\ref{eq:AC1}), (\ref{eq:AC2}) and (\ref{eq:AC3}) into
(\ref{eq:AC}), we get
\begin{equation}
g^{ab}u^{c}\sigma_{ae}\sigma_{bf}\nabla_{c}\nabla^{e}u^{f}=-\theta_{ab}\theta^{ba}+\boldsymbol{\mathfrak{D}}\cdot\boldsymbol{\alpha}+\aleph\left(\Theta-\aleph\right)+\left(\nabla_{\boldsymbol{u}}n_{b}\right)\nabla_{\boldsymbol{n}}u^{b}-R_{ab}u^{a}u^{b}+R_{acbd}n^{a}n^{b}u^{c}u^{d}.\label{eq:ARTerm}
\end{equation}
Now, inserting (\ref{eq:AAlphaSquaredTerm}), (\ref{eq:AThetaabTerm})
and (\ref{eq:ARTerm}) into (\ref{eq:AUnsimplified}), the terms involving
the derivatives of $\boldsymbol{n}$ along $\boldsymbol{u}$ as well
as the square of the normal component of the acceleration cancel,
leaving the Raychaudhuri equation in the form
\begin{equation}
\frac{{\rm d}\theta}{{\rm d}\tau}=-\theta_{ab}\theta^{ba}+\left(\boldsymbol{\alpha}+\boldsymbol{\mathfrak{D}}\right)\cdot\boldsymbol{\alpha}+\aleph\Theta+u^{a}u_{c}\Theta_{ab}\Theta^{bc}-R_{ab}u^{a}u^{b}+R_{acbd}n^{a}n^{b}u^{c}u^{d}.\label{eq:ASimplified}
\end{equation}
To get from this to (\ref{eq:Raychaudhuri}), we can check from the
definition of quasilocal momentum $\mathcal{P}^{a}$ that
\begin{equation}
\mathcal{P}^{a}\mathcal{P}_{a}=u^{a}u_{c}\Theta_{ab}\Theta^{bc}+\left(u^{a}u^{b}\Theta_{ab}\right)^{2}.
\end{equation}
Also, it can be checked (via the Leibnitz rule and orthogonality properties)
that $u^{a}u^{b}\Theta_{ab}=-\aleph$. Using $\Theta=\kappa+\aleph$,
this entails that
\begin{equation}
\boldsymbol{\mathcal{P}}^{2}+\kappa\aleph=u^{a}u_{c}\Theta_{ab}\Theta^{bc}+\Theta\aleph.
\end{equation}
Inserting this into (\ref{eq:ASimplified}) we finally get 
\begin{equation}
\frac{{\rm d}\theta}{{\rm d}\tau}=-\theta_{ab}\theta^{ba}-R_{ab}u^{a}u^{b}+\boldsymbol{\mathfrak{D}}\cdot\boldsymbol{\alpha}+\boldsymbol{\alpha}^{2}+\boldsymbol{\mathcal{P}}^{2}+\aleph\kappa+R_{acbd}n^{a}n^{b}u^{c}u^{d}.\label{eq:RaychaudhuriCurlyD}
\end{equation}
Making use of $0=D_{a}\gamma^{ab}=\bm{a}\cdot\bm{u}=\sigma^{ab}n_{b}=D_{a}n_{b}$
and $a_{b}=\alpha_{b}+\aleph n_{b}$, we remark that
\begin{align}
\bm{D}\cdot\bm{\alpha} & =D_{a}\left(\sigma^{ab}a_{b}\right)\\
 & =\left(D_{a}\sigma^{ab}\right)a_{b}+\sigma^{ab}\left(D_{a}a_{b}\right)\\
 & =\left(D_{a}\left(\gamma^{ab}+u^{a}u^{b}\right)\right)a_{b}+\sigma^{ab}\left(D_{a}\left(\alpha_{b}+\aleph n_{b}\right)\right)\\
 & =\left(D_{\bm{u}}u^{b}\right)a_{b}+\mathfrak{D}^{b}\alpha_{b}\\
 & =\gamma^{ba}a_{a}a_{b}+\boldsymbol{\mathfrak{D}}\cdot\boldsymbol{\alpha}\\
 & =\left(\sigma^{ab}-u^{a}u^{b}\right)a_{a}a_{b}+\boldsymbol{\mathfrak{D}}\cdot\boldsymbol{\alpha}\\
 & =\bm{\alpha}\cdot\bm{a}+\boldsymbol{\mathfrak{D}}\cdot\boldsymbol{\alpha}\\
 & =\bm{\alpha}\cdot\bm{\alpha}+\boldsymbol{\mathfrak{D}}\cdot\boldsymbol{\alpha},
\end{align}
so (\ref{eq:RaychaudhuriCurlyD}) is precisely (\ref{eq:Raychaudhuri}).

\renewcommand{\thesection}{B}
\section{The Raychaudhuri Equation for the Kerr GQF}

\label{sec:B} 

Let us analyze the Raychaudhuri GQF equation (\ref{eq:RaychaudhuriGeoid})
with (\ref{eq:KerrGQFsoln}). For this solution, the Ricci ternsor
term vanishes, and so the equation involves only four terms: the acceleration
term $\aleph\kappa$, the tidal term $R_{acbd}n^{a}n^{b}u^{c}u^{d}$,
the quasilocal momentum squared $\boldsymbol{\mathcal{P}}^{2}$, and
the expansion term $-\theta_{ab}\theta^{ba}$. We compute all of them
explicitly, and find that, as expected, they sum exactly to zero,
i.e. our solution indeed satisfies the Raychaudhuri GQF equation.
Their full expressions are rather cumbersome and we refrain from writing
them out in full here, but it may be instructive to look at their
expansions in the angular momentum parameter $a$:

\begin{align}
\aleph\kappa & =\frac{2M}{F^{3}\left(r\right)}-M\left\{ \frac{2\left[3F\left(r\right)-2M\right]\cos^{2}\theta}{F^{5}\left(r\right)\left[F\left(r\right)-2M\right]}+\frac{\left[7M-3F\left(r\right)\right]\sin^{2}\theta}{F^{4}\left(r\right)\left[F\left(r\right)-2M\right]^{2}}\right\} a^{2}+\ldots,\label{eq:KerrR1}\\
R_{acbd}n^{a}n^{b}u^{c}u^{d} & =-\frac{2M}{F^{3}\left(r\right)}+3M\left\{ \frac{2\cos^{2}\theta}{F^{5}\left(r\right)}-\frac{\sin^{2}\theta}{F^{4}\left(r\right)\left[F\left(r\right)-2M\right]}\right\} a^{2}+\ldots,\label{eq:KerrR2}\\
\boldsymbol{\mathcal{P}}^{2} & =\frac{M^{2}\sin^{2}\theta}{F^{4}\left(r\right)\left[F\left(r\right)-2M\right]^{2}}a^{2}+\ldots,\label{eq:KerrR3}\\
-\theta_{ab}\theta^{ba} & =8\frac{M^{2}\cos^{2}\theta}{F^{5}\left(r\right)\left[F\left(r\right)-2M\right]}a^{2}+\ldots,\label{eq:KerrR4}
\end{align}
where $\ldots$ refers to $\mathcal{O}\left(a^{4}\right)$ terms.

As $a\rightarrow0$, we see that we recover the terms we found in
the Schwarzschild GQF.

\begin{figure}
\noindent \begin{centering}
\includegraphics[scale=0.4]{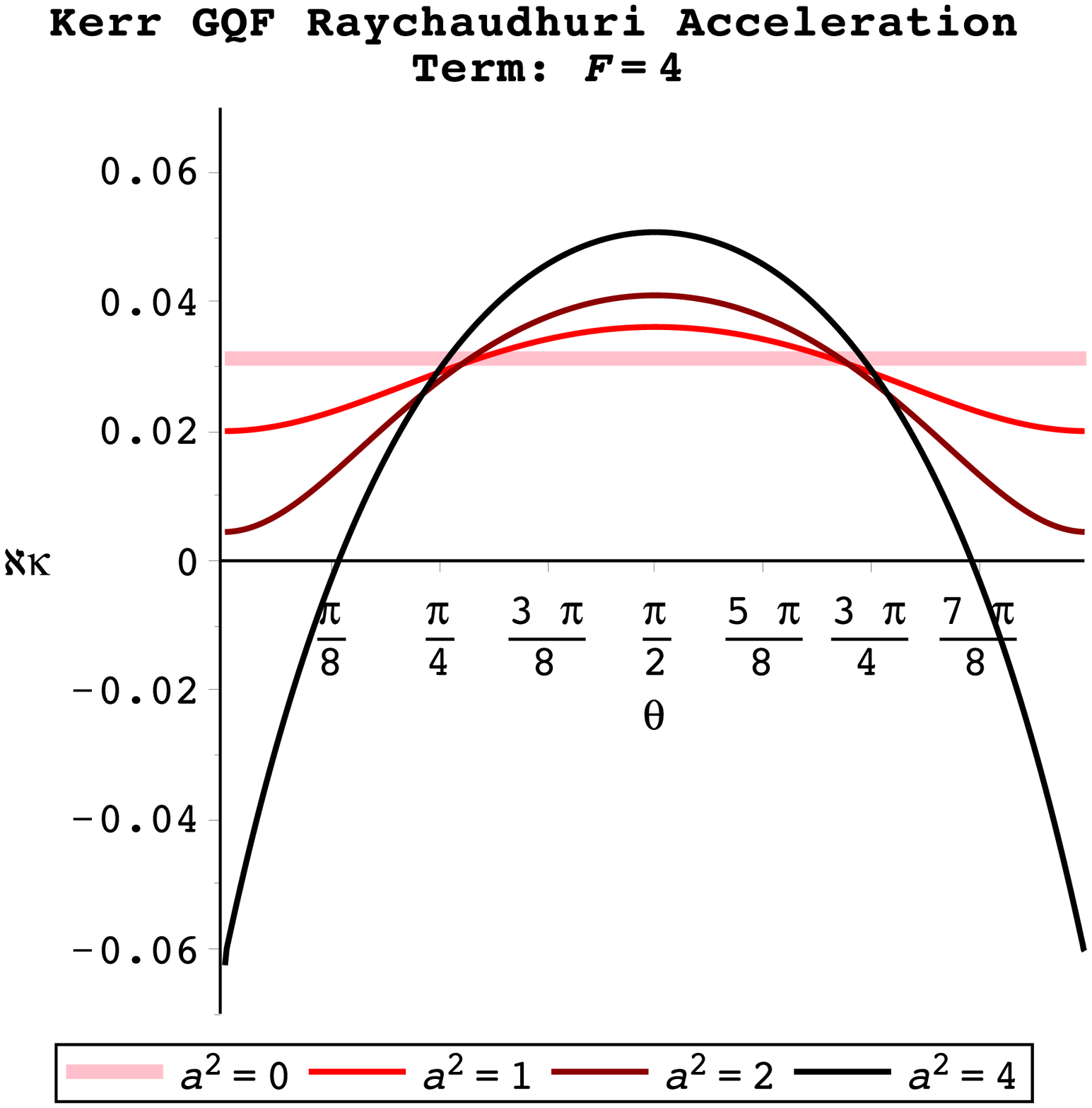} \includegraphics[scale=0.4]{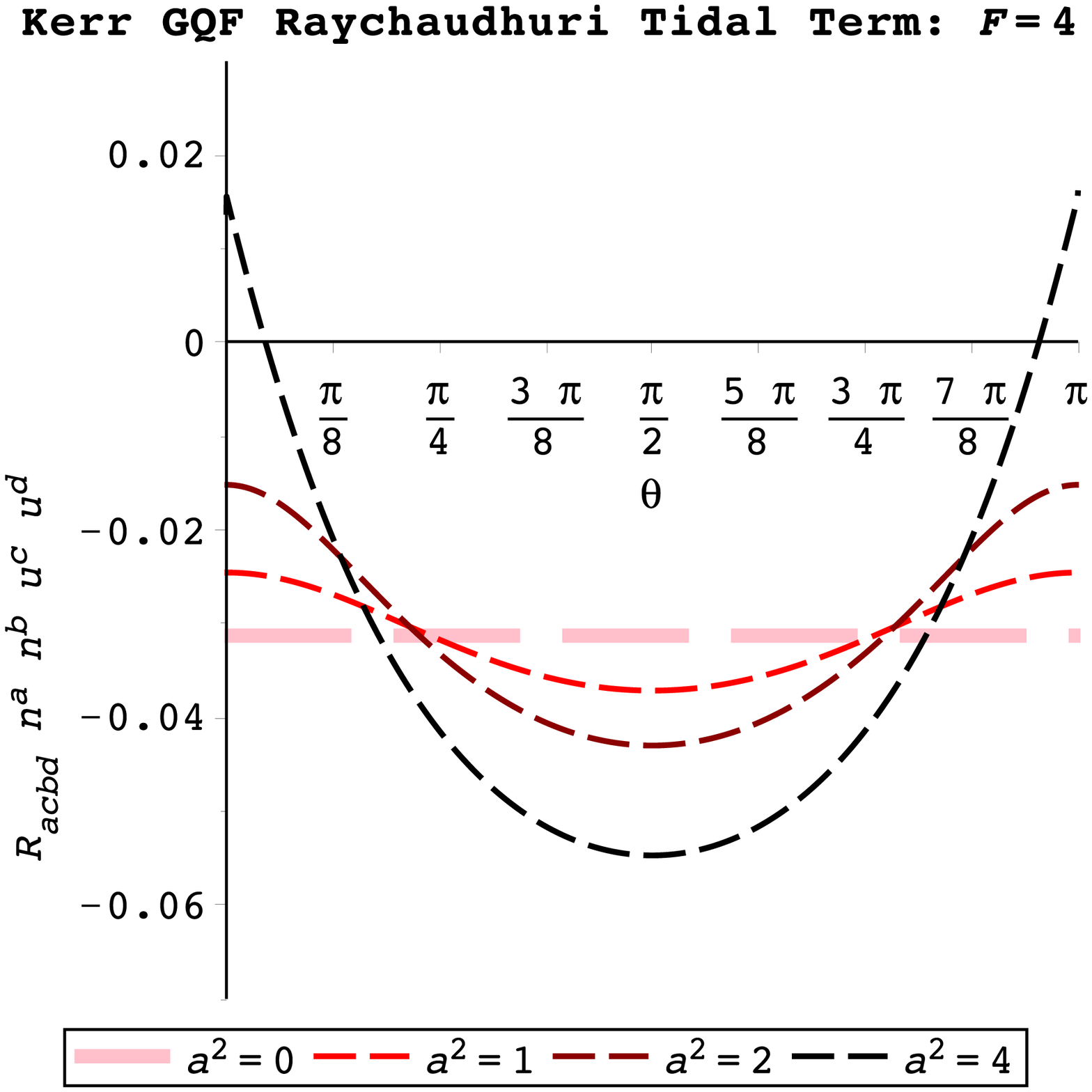} 
\par\end{centering}

~

\noindent \begin{centering}
\includegraphics[scale=0.4]{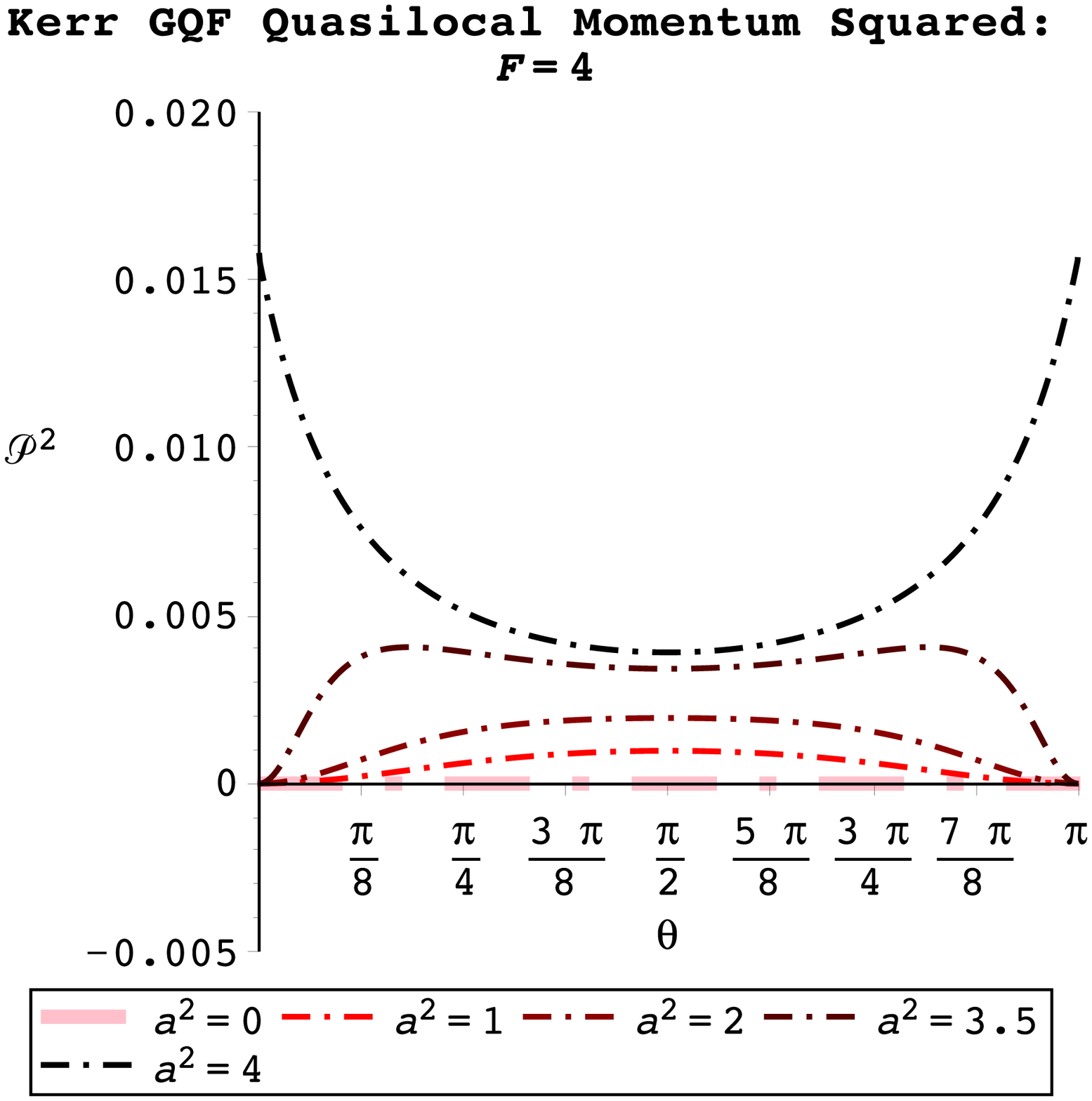} \includegraphics[scale=0.4]{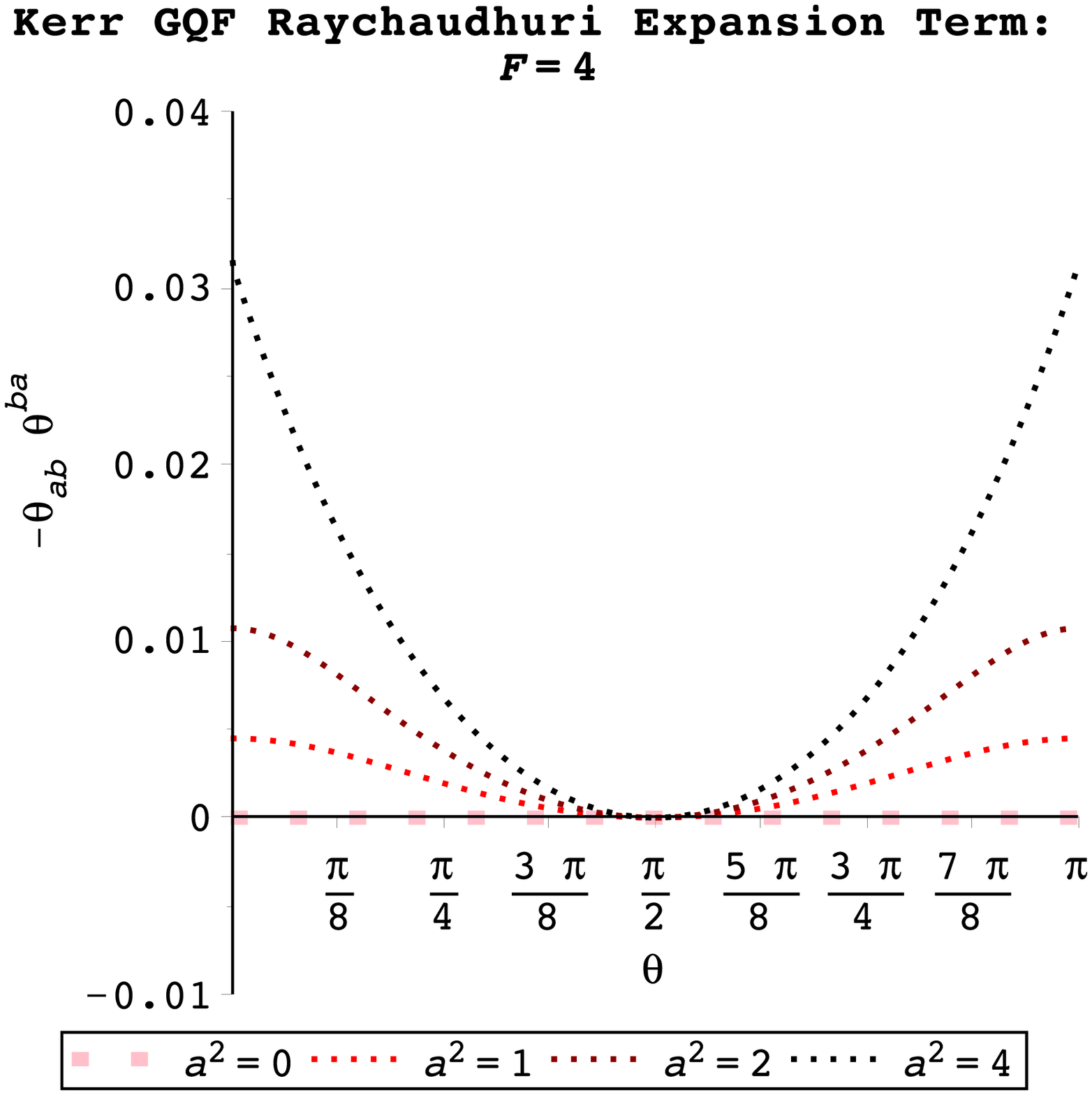} 
\par\end{centering}

\protect\protect\caption{\label{fig:2d_plot_kerr_F=00003D00003D4_RGQF_separately}The Raychaudhuri
terms for the Kerr GQF with $F\left(r\right)=r=4M$ in units of $M=1$,
plotted separately for different values of the angular momentum parameter
$a$. (Increasing colour darkness corresponds to increasing $a$.)}
\end{figure}

\begin{figure}
\noindent \begin{centering}
\includegraphics[scale=0.4]{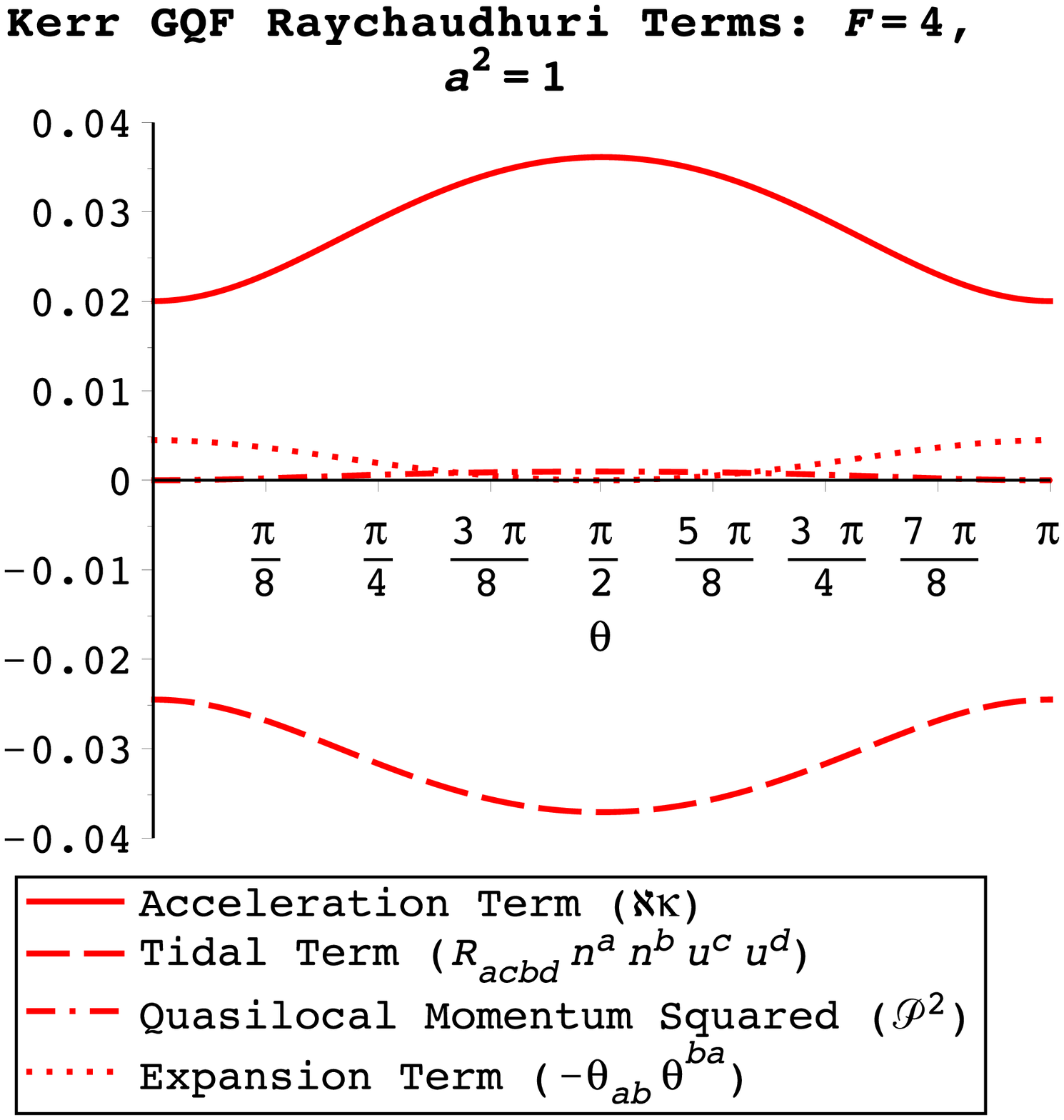} \includegraphics[scale=0.4]{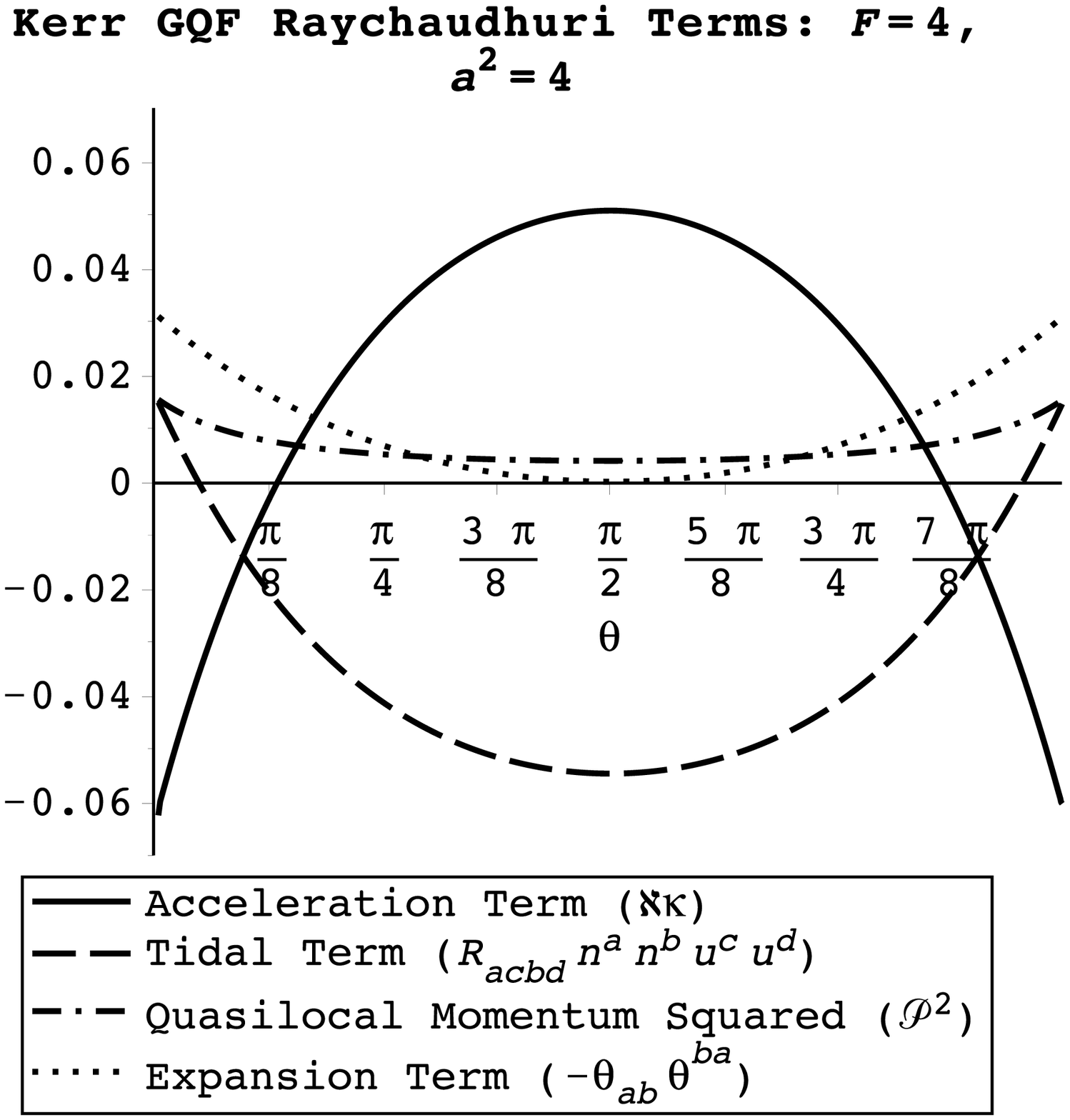} 
\par\end{centering}

\protect\protect\caption{\label{fig:2d_plot_kerr_F=00003D00003D4_RGQF_together}The Raychaudhuri
terms for the Kerr GQF with $F\left(r\right)=r=4M$ in units of $M=1$,
plotted together for angular momentum parameter $a^{2}=1$ (left)
and $a^{2}=4$ (right).}
\end{figure}

Setting $F\left(r\right)=r$, we plot the (exact) terms at $r=4M$
in units of $M=1$, for different $a^{2}$ values -- one by one in
Figure \ref{fig:2d_plot_kerr_F=00003D00003D4_RGQF_separately}, and
against each other in Figure \ref{fig:2d_plot_kerr_F=00003D00003D4_RGQF_together}.

\bibliography{references_v3}

\begin{thebibliography}{10}

\bibitem{GRACE}
B.~D. Tapley, S.~Bettadpur, M.~Watkins, and C.~Reigber.
\newblock The gravity recovery and climate experiment: Mission overview and
  early results.
\newblock {\em Geophys. Res. Lett.}, 31:L09607, May 2004.

\bibitem{GOCE}
Roland Pail~et al.
\newblock First goce gravity field models derived by three different
  approaches.
\newblock {\em Journal of Geodesy}, 85(11):819--843, 2011.

\bibitem{EGM98}
Dru~A. Smith.
\newblock There is no such thing as ``the" egm96 geoid: Subtle points on the
  use of a global geopotential model.
\newblock {\em IGeS Bulletin No. 8, International Geoid Service, Milan, Italy},
  pages 17--28, 1998.

\bibitem{Moritz}
Bernhard Hofmann-Wellenhof and Helmut Moritz.
\newblock {\em Physical Geodesy}.
\newblock Springer-Verlag, Vienna, 2006.

\bibitem{NOAA}
Nicole Kinsman.
\newblock Personal communication. 09/11/2015. National Oceanic and Atmospheric
  Administration/National Ocean Service/National Geodetic Survey. Silver
  Spring, Maryland.

\bibitem{GPS}
D.G. Milbert and D.A. Smith.
\newblock Converting gps height into navd 88 elevation with the geoid96 geoid
  height model.
\newblock {\em GIS/LIS '96 Annual Conference and Exposition. American Congress
  on Surveying and Mapping, Washington, DC}, pages 681--692, 1996.

\bibitem{Kopeikin:2014afa}
Sergei Kopeikin, Elena Mazurova, and Alexander Karpik.
\newblock {Towards an exact relativistic theory of Earth's geoid undulation}.
\newblock {\em Phys.Lett.}, A379:1555--1562, 2015.

\bibitem{Epp:2008kk}
Richard~J. Epp, Robert~B. Mann, and Paul~L. McGrath.
\newblock {Rigid motion revisited: Rigid quasilocal frames}.
\newblock {\em Class.Quant.Grav.}, 26:035015, 2009.

\bibitem{Epp:2013xza}
Richard~J. Epp, Robert~B. Mann, and Paul~L. McGrath.
\newblock {On the Existence and Utility of Rigid Quasilocal Frames}.
\newblock 2013.

\bibitem{McGrath:2012db}
Paul~L. McGrath, Richard~J. Epp, and Robert~B. Mann.
\newblock {Quasilocal Conservation Laws: Why We Need Them}.
\newblock {\em Class.Quant.Grav.}, 29:215012, 2012.

\bibitem{Epp:2013hua}
Richard~J. Epp, Paul~L. McGrath, and Robert~B. Mann.
\newblock {Momentum in General Relativity: Local versus Quasilocal Conservation
  Laws}.
\newblock {\em Class.Quant.Grav.}, 30:195019, 2013.

\bibitem{McGrath:2013pea}
Paul~L. McGrath, Melanie Chanona, Richard~J. Epp, Michael~J. Koop, and
  Robert~B. Mann.
\newblock {Post-Newtonian Conservation Laws in Rigid Quasilocal Frames}.
\newblock {\em Class.Quant.Grav.}, 31:095006, 2014.

\bibitem{Poisson}
Eric Poisson.
\newblock {\em A Relativist's Toolkit: The Mathematics of Black-Hole
  Mechanics}.
\newblock Cambridge University Press, Cambridge, 2007.

\bibitem{Wald}
Robert~M. Wald.
\newblock {\em General Relativity}.
\newblock University Of Chicago Press, Chicago, 1973.

\bibitem{Brown-York:1993}
J.~David Brown and James~W. York.
\newblock Quasilocal energy and conserved charges derived from the
  gravitational action.
\newblock {\em Phys. Rev. D}, 47:1407--1419, Feb 1993.

\bibitem{Brown:2000dz}
J.~David Brown, S.R. Lau, and Jr. York, James~W.
\newblock {Action and energy of the gravitational field}.
\newblock 2000.

\bibitem{Breton}
N.~Bret\'on, A.~A. Garc\'{i}a, V.~S. Manko, and T.~E. Denisova.
\newblock Arbitrarily deformed kerr-newman black hole in an external
  gravitational field.
\newblock {\em Phys. Rev. D}, 57:3382--3388, Mar 1998.

\bibitem{Adler}
Ronald Adler, Maurice Bazin, and Menahem Schiffer.
\newblock {\em Introduction to General Relativity}.
\newblock McGraw-Hill, New York, 1965.

\bibitem{Frutos-Alfaro:2014usa}
Francisco Frutos-Alfaro.
\newblock {Perturbation of the Kerr Metric}.
\newblock arXiv:1401.0866.

\bibitem{Kopeikin:2015}
Sergei Kopeikin, Wenbiao Han, and Elena Mazurova.
\newblock {Post-Newtonian reference-ellipsoid for relativistic geodesy}.
\newblock arXiv:1510.03131.

\end{thebibliography}

\end{document}